\documentclass[
 nofootinbib,
 a4paper,
 aps,
 prx,
 twocolumn,
 amsmath,
 superscriptaddress,
]{revtex4-2}
\usepackage{graphicx}
\usepackage{amsfonts}
\usepackage[dvipsnames]{xcolor}
\usepackage[colorlinks=true,citecolor=blue,linkcolor=blue,urlcolor=blue]{hyperref}
\pdfoutput=1
\usepackage{amsmath,amssymb,mathtools}
\usepackage{physics}
\usepackage{bm}
\usepackage[dvipsnames]{xcolor}
\usepackage[USenglish]{babel}
\usepackage[capitalize]{cleveref}
\usepackage{appendix}

\usepackage[utf8]{inputenc}
\usepackage[T1]{fontenc}

\usepackage[activate={true,nocompatibility},final,tracking=true,kerning=true]{microtype}

\usepackage{ulem}

\newcommand{\beq}{\begin{equation}}
\newcommand{\eeq}{\end{equation}}

\AtBeginEnvironment{appendices}{\crefalias{section}{appendix}}
\graphicspath{{./}{./figures/}}

\begin{document}

\title{Measuring temporal entropies in experiments}
\author{Aleix Bou-Comas}
\affiliation{Institute of Fundamental Physics IFF-CSIC, Calle Serrano 113b, Madrid 28006, Spain}

\author{Carlos Ramos Marimón}
\affiliation{Departament de Física Quàntica i Astrofísica and Institut de Ciències del Cosmos (ICCUB), Universitat de Barcelona, Martí i Franquès 1, 08028 Barcelona, Spain}

\author{Jan T.\ Schneider}
\affiliation{Institute of Fundamental Physics IFF-CSIC, Calle Serrano 113b, Madrid 28006, Spain}

\author{Stefano Carignano}
\affiliation{Barcelona Supercomputing Center, 08034 Barcelona, Spain}

\author{Luca Tagliacozzo}
\email{luca.tagliacozzo@iff.csic.es}
\affiliation{Institute of Fundamental Physics IFF-CSIC, Calle Serrano 113b, Madrid 28006, Spain}

\begin{abstract}
We propose a novel experimental protocol to measure generalized temporal entropies in many-body quantum systems. Our approach involves using local operators as probes to characterize the out-of-equilibrium dynamics induced by a geometric double quench on a replicated system. Such protocol mimics the path-integral on the corresponding Riemann surface encoding  generalized temporal entanglement. We present the results of tensor network simulations of one-dimensional systems which validate the protocol and demonstrate the experimental feasibility of measuring generalized temporal entropies, and we outline the experimental requirements for implementing these quenches using state-of-the-art quantum simulators.
Therefore, our results provide a physical interpretation of the meaning of generalized temporal entropies. Furthermore, they  reveal that the dynamics induced on two replicas of the Ising model in a transverse field differ qualitatively from the ones of its non-integrable extension, suggesting that generalized temporal entropies can be used as a tool for identifying different dynamical classes in quantum systems.
\end{abstract}
\maketitle

\section{Introduction}
The behavior of many-body quantum systems underpins everything from the materials we use, to the vacuum of quantum field theories like Quantum Electrodynamics. In such systems often emerge collective behavior, where elementary excitations behave fundamentally differently from individual constituents. For instance, electrons can pair to form bosonic Cooper pairs in superconductors, or quarks bind to form color-neutral hadrons in nuclear matter. These emergent phenomena are closely tied to strong correlations and are characterized by distinct patterns of entanglement—a feature that remains at the heart of modern condensed matter theory.

Entanglement, which describes the non-local correlations between parts of a system~\cite{amico2008,laflorencie2016,zeng2018}, has become a vital tool for understanding a range of phenomena, from quantum phase transitions~\cite{callan1994,vidal2003a,calabrese_2004} to the distinction between exotic phases of matter~\cite{kitaev2006,levin2006}. Moreover, it plays a crucial role in out-of-equilibrium dynamics, helping to reveal the degree of ergodicity in these systems~\cite{calabrese_2005,bardarson2012,serbyn2013}. Importantly, recent advances allow measuring entanglement and related properties in experiments, further confirming its prominent role in both theoretical and experimental physics~\cite{cardy2011,abanin2012,daley2012,islam2015,kaufman2016,hauke_2016,brydges2019,huang2020,elben2020}.
Traditional entanglement measures typically capture the spatial correlations between different parts of a system at a single moment in time. However, a new avenue of research investigates the entanglement across different times, a concept we refer to as \emph{temporal entanglement}~\cite{leifer2006,leifer2013,parzygnat2023}. Understanding the nature of these correlations requires going beyond the usual Hamiltonian dynamics and adopting the path integral formulation of quantum mechanics, where time and space are treated on equal footing.
\begin{figure}[t]
  \includegraphics[width=\linewidth]{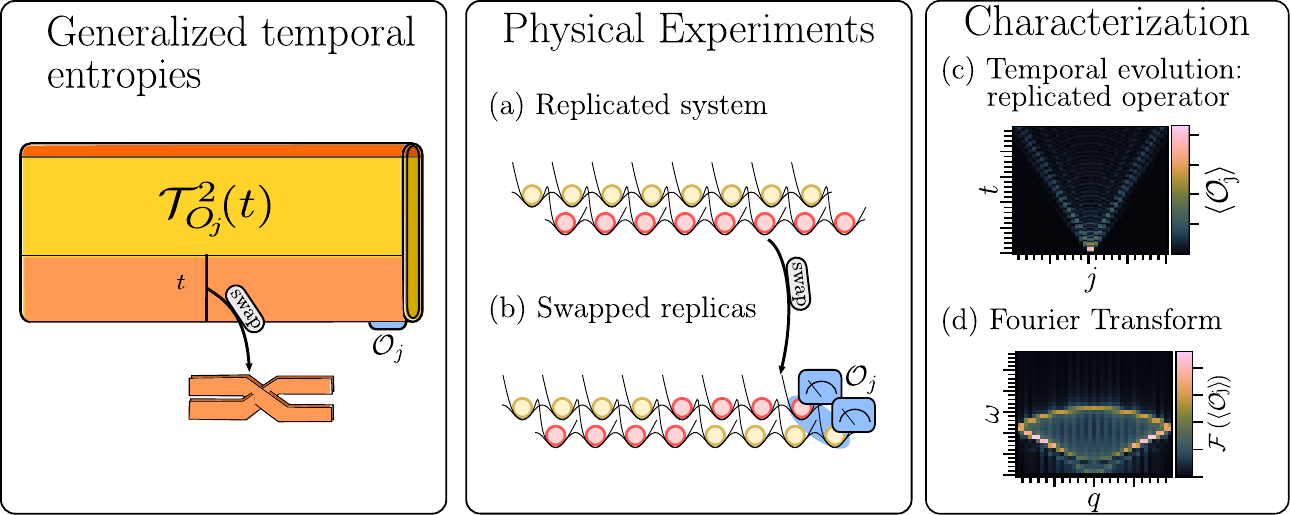}
  \caption{\label{fig:quench_fig}
  The path integral formulation leading to the generalized temporal purities (left) can be mapped explicitly to a double-quench experimental
  setup (center). We exemplify the idea by using atoms trapped in optical lattices (a): we start with two replicas with a deep super-lattice in the y-direction to keep them separated. The two copies start evolving separately under the same
Hamiltonian. At a given time, dynamics is frozen by raising the potential in the x-direction and a swap is performed by  lowering, in the region we want to swap, the lattice in the y-direction for the time needed to obtain a full transfer of the atoms between the two wells. (b) shows the configuration after the swap, leading to the subsequent independent evolution of the partially swapped replicas.  We finally measure any local observable jointly on the two copies.
  By repeating several experiments, varying both the time after the swap and the position where the observable is measured, we can access a full space-time picture (right). Its Fourier transforms (d) reveals spectral properties of the system defined on a Riemann surface.
}
%

%
\end{figure}

In this formulation, the Feynman--Vernon influence functional~\cite{feynman1963} serves as a key tool, capturing the history of a system by isolating the interactions of a single constituent across different times. By treating this functional as a vector in a multi-time Hilbert space, we can compute reduced density matrices and their associated temporal entanglement. Notice that such quantity is strictly speaking the operator entanglement of the influence functional and thus mixes quantum and classical correlations. While classical simulations of dynamics that generate limited temporal entanglement are feasible, the full implications of temporal entanglement for many-body dynamics remain an open question~\cite{banuls2009,muller-hermes2012,hastings2015,lerose2021,sonner2021,giudice2022,carignano2024a,guo2024a,yao2024a,foligno2023a,yao2024a}.
Despite these advances, the study of temporal correlations, and possible measures of temporal entanglement in many-body quantum systems is still in its early stages.

In field theories, a different concept of temporal entanglement has been explored through reduced transition matrices, which arise by cutting the path integral along time-like paths~\cite{nakata2021,doi2023a,doi2023b,narayan2023,narayan2023b,heller2024}. These matrices encode complex-valued generalized entropies~\cite{mollabashi2021,nakata2021,murciano2022} that carry geometric information in holographic field theories. A related notion of generalized temporal entropies was introduced by some of the authors in~\cite{carignano2024a}, arising from a time-like cut of the path integral encoding the time-dependent expectation value of a local operator.
Generalized temporal entropies coincide with studying path integrals on replicated Riemann surfaces with a branch cut connecting the different sheets lying on the chosen time-like path.  In this work, we shall focus on this particular definition, and show that these mathematical objects coincide with the outcome of a pump-probe experiment, where a geometric quench excites the system and the operators probes the resulting evolution.

This has several important consequences:  
 First, it implies  that \textit{generalized temporal Rényi entropies can be accessed experimentally}, by actually performing such a pump probe experiment. Second, that \textit{they are real-valued},  as they correspond to the expectation value of Hermitian operators on replicated states generated by a two-step quench protocol. Third, that \textit{they are ultraviolet finite}, in the limit of a continuous evolution of a lattice system. Finally, we show that generalized temporal entropies provide a powerful tool for characterizing the out-of-equilibrium dynamics: through extensive tensor network simulations of the transverse field Ising model ~\cite{banuls2009,hastings2015,carignano2024a}, we observe distinct behaviors of temporal entropies in its integrable and non-integrable regimes.

 The simplest experimental realization of our proposed protocol is sketched in \cref{fig:quench_fig}. We
 begin the experiment with two decoupled copies of the same system, corresponding to the two sheets of the path integral's Riemann surface.  The two copies start evolving independently under the same Hamiltonian. At a given instant of time, corresponding to our temporal bipartition, a slice of the two systems is then swapped, after which the two copies keep evolving separately. We finally probe the system by measuring the same local observable on the two copies.
 This gives access to a full space-time picture, allowing to investigate spectral properties of the system defined on a Riemann surface.
 
 This paper is organized as follows: we begin in Sec.~\ref{sec:gen_reny_entro}
 by defining the generalized temporal entropies via a path integral formulation on multi-sheeted Riemann surfaces, and show in Sec.~\ref{sec:double_quench} how can they be related to a double quench measurement, exposing some of their properties. We investigate in Sec.~\ref{sec:footprint-integrability} how the spatio-temporal features of these entropies are affected by the properties of the underlying models, and propose in Sec.~\ref{sec:exp} concrete experimental setups for their measurement. We finally point out implications and future directions of this work in Sec.~\ref{sec:conclusions}.

\section{Definition of temporal entropies}\label{sec:gen_reny_entro}

\begin{figure}
  \centering
  \includegraphics[width=0.95\linewidth]{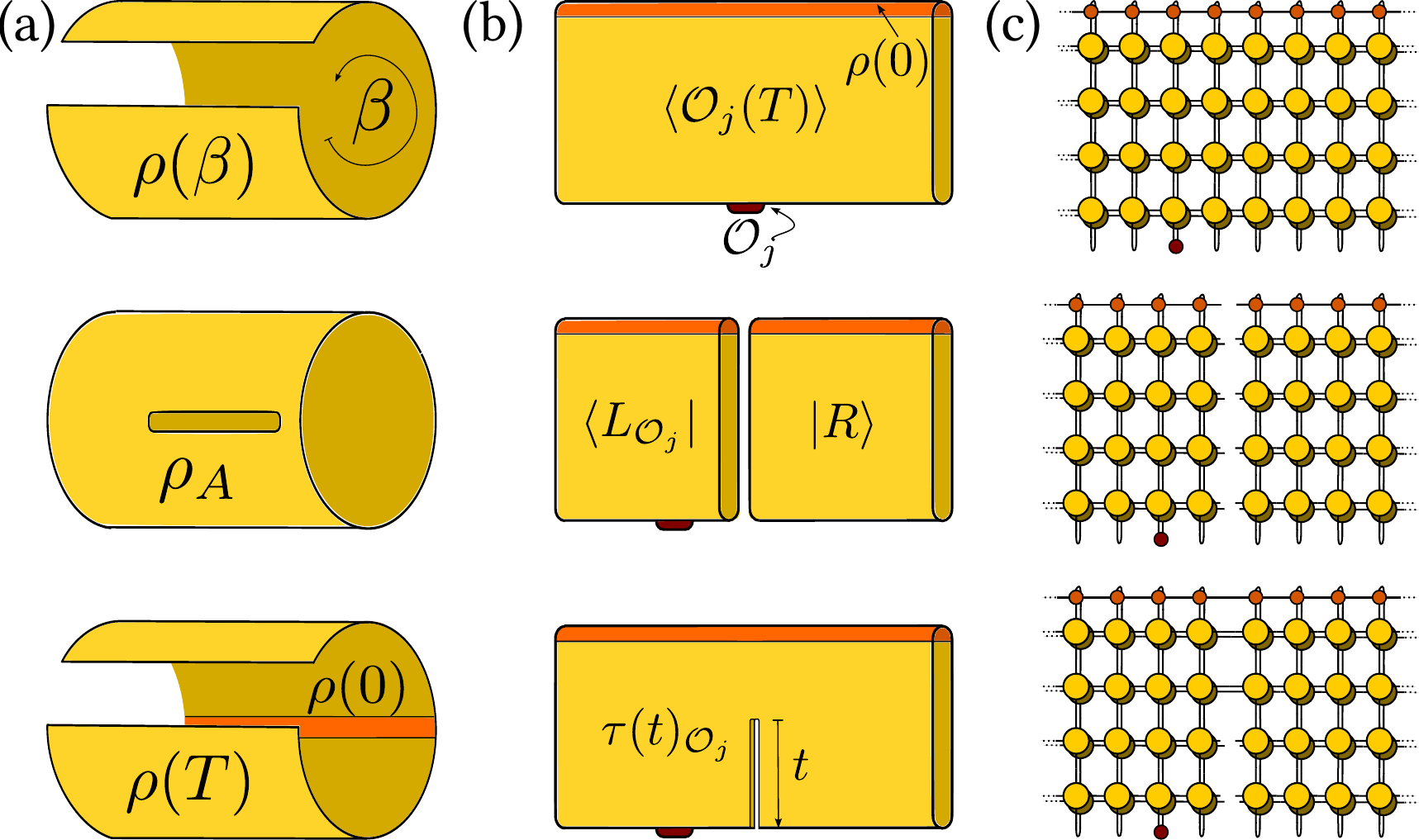}
  \caption{\label{fig:path_integral}%
    Surfaces representing different path integrals. These can be interpreted either as a target space-time of the defining field-theory path integrals, or as idealized representations of the contraction of 2D networks whose elementary tensors encode such states. (a) From top to bottom, we sketch the equilibrium state $\rho(\beta)$, the reduced density matrix $\rho_A = \tr_{\neg A}(\rho)$, and temporal evolution of a 1D many-body system $\rho(T)$. In particular, the standard reduced density matrix of a region $A$, represented in the middle, is obtained by a path integral with a horizontal cut along $A$.
    (b) From top to bottom, we represent the time-dependent expectation value of an operator. Here the cylinder is flattened, times runs vertically and space horizontally. In the middle, we show the two temporal vectors $\bra*{L_{O_j}}$, $\ket{R}$, obtained by cutting the original path integral in the time direction (vertical). Their overlap encodes the time-dependent expectation value, while their partial contraction gives rise to the reduced transition matrices $\tau_{O_j}(t)$ of \cref{eq:red_tm}. The latter are represented at the bottom as the path integral with a partial temporal cut. From $\tau_{O_j}(t)$, one can extract the generalized temporal entropies using \cref{eq:gen_reny_entro}.
    (c) The same quantities as in column (b) but with an explicit tensor-network representation.
  }
\end{figure}

We focus on systems defined on a one dimensional lattice, which can be viewed either as regularizations of 1D field theories or as many-body quantum systems, described by a local Hamiltonian, $H = \sum_i h_i$, where $i$ represents lattice's sites and $h_i$ an operator acting on $i$ and its neighbors.
At equilibrium, the state of the system is described by the density matrix
\begin{equation}
 \rho(\beta) = \frac{\exp(-\beta H)}{\mathcal{Z}} \,,\quad {\mathcal{Z}} = \Tr[ \exp(-\beta H)] \,, \,\label{eq:path_integral}
\end{equation}
where $\beta$ is the inverse temperature.
In our computations, we discretize the (Euclidean) time in elementary time steps $\delta t$. The total Euclidean time evolution is then obtained after $N_{\beta} = \beta / \delta t$ infinitesimal steps. {Leveraging the locality of the Hamiltonian, each infinitesimal step can be approximated to arbitrary order in $\delta t$ using, for example, a Suzuki-Trotter approximation~\cite{vidal2003}}. { This generates a matrix product operator (MPO) approximation of the infinitesimal step that involves the contraction of $N$ elementary tensors~\cite{schollwock2011,paeckel2019}.}
As a result, Eq.~\eqref{eq:path_integral} for a system of $N$ constituents can be encoded as the contraction of a 2D space-time tensor network with $N\times N_{\beta}$ tensors.
In \Cref{fig:path_integral}, we represent this 2D tensor network as a surface, allowing for a connection to path integral formulations used in holographic field theories\footnote{This relation can be made quantitative for example close to the continuum limit of a lattice model, see e.g.,~\cite{carignano2024}}.
The expectation value of a local operator on a site $j$ computed as $\langle \mathcal{O}_j\rangle = \tr(\mathcal{O}_j \rho(\beta))$, thus the surface is a cylinder with circumference $\beta$ and length $N$, as depicted in \cref{fig:path_integral}.
In absence of degeneracy, the ground state is obtained by considering the zero temperature limit
$\ketbra{\Omega}=\lim_{\beta\to \infty} \rho(\beta)$, which results in a cylinder with an infinitely large circumference.

We can compute the entanglement of a region $A$ of the system with its complement $B$ by tracing the degrees of freedom in $B$, yielding the reduced density matrix for the region $A$, $\rho_A=\tr_B \rho(\beta)$. In our diagrammatic illustration, the partial trace implies that the extremes of the cylinder are only partially glued along $B$, while remaining open along $A$, as shown in the middle of the left panel of \cref{fig:path_integral},~\cite{callan1994,calabrese_2004}.

A similar construction holds for out-of-equilibrium states $\rho(T)$, with $T$ real time,
generated after a quench from $H$ to some $H'$ governing the dynamics:
\begin{equation}
 \rho(T) = U(T) \rho(0)U ^ \dagger(T) \,\label{eq:quench}\,,
\end{equation}
where  $H$ defines the initial state $\rho(0)$  through \cref{eq:path_integral} and we defined $U(T)= \exp(-iH'T)$. 
By repeating the same steps and defining $N_T= T/\delta t$, in this case the expectation value $\langle O_j(T)\rangle \equiv \tr(O_j \rho(T) )$ is encoded in a two-dimensional tensor network containing $2N_T\times N$ tensors.
The factor $2$ arises from  the presence of both $U$ and $U^\dagger$ each generating half of the total cylinder, made by a forward and a backward contour glued together by the initial state.
The trace is then used to compute the expectation value of the local operator.
Upon taking such trace without the local operator, the cylinder would collapse, i.e., trivialize to the trace of the initial state since $U U^\dagger = \mathbb{I}$.
The insertion of $O_j$ is therefore sufficient to avoid such cancellation, and it unveils the 2D space-time nature of the tensor network, as shown on the upper sketch of panel (b) in \cref{fig:path_integral}.

In order to define generalized temporal entropies, we proceed analogously to what we do in the case of standard entropies and
 define a temporal cut at spatial position $r$ on the path integral defining $\expval*{O_j (T)}$. Without loss of generality, we position the cut on the right of the location of the operator.
As a result, the expectation value $\expval*{O_j (T)}$ can be written as the overlap of two temporal states $\bra*{{L}_{r,O_j}}$ and $\ket{{R}_{r}}$\footnote{Whenever $r=N/2$, with $N$ the spatial extent of the system, we will omit it from the definition of the temporal states.}, as illustrated in the middle panel of \cref{fig:path_integral}(b):
\begin{equation}
\langle O_j(T)\rangle \equiv \bra{{L}_{O_j}}\ket{R} \, . \label{eq:def_L_R}
\end{equation}

These two boundary states live in the Hilbert space spanned by the horizontal legs of the MPO crossing the cut, connecting the first $r\times N_{T}$ space-time tensors to the left of the vertical cut with the remaining $(N-r)\times N_T$ space-time tensors to the right, as shown in the central panel of \cref{fig:path_integral} (c).
Thus, they encode multi-time vectors each with $2N_T$ ($N_T$ from $U$ and $N_T$ from the $U^\dagger$) legs~\cite{carignano2024a}. They are mixed temporal states, the left and right influence functionals,  having support on both the forward and backward sheet.

To better understand this point, we can realize that when initial state is pure,  $\bra*{{L}_{O_j}}$ and $\ket{R}$ are the reduced density matrices of the temporal degrees of freedom of the process tensor obtained by cutting in two halves the evolution operator $U(T)$ applied to the initial state $\ket{\psi_0}$ (see~\cite{cerezo-roquebrún2025} for a detailed review of the properties and connections between the different objects we are introducing here). In the following we will vectorize these mixed states, and thus the entropies we will introduce should be considered as operator entropies in the spirit of the operator entanglement~\cite{zanardi2001,dubail2017}.

We can perform partial overlaps by contracting, for example, only the first $2(N_T-N_t)$ legs of these states, leaving the remaining $2N_t$ legs open as shown in the lower panel of \cref{fig:path_integral}(c). This generates a tensor network realization of the \emph{temporal reduced transition matrices}~\cite{nakata2021,murciano2022,doi2023a,doi2023b,narayan2023,narayan2023b,carignano2024,carignano2024a}, analogous to the standard reduced density matrix,
\begin{equation}
 {\tau_{O_j}(t)}= \frac{\tr_{T-t}\ketbra{R}{L_{O_j}}}{\braket{L_{O_j}}{R}}\label{eq:red_tm} \,.
\end{equation}
Since $\ket{R}\neq \ket{L}$ in general, the partial overlap in \cref{eq:red_tm} gives rise to matrices with complex-valued entries and complex-valued spectra (when they exist). Nonetheless, we can define \emph{generalized temporal Rényi entropies} based on the traces of powers of these reduced transition matrices,
\begin{equation}
\mathcal{T}^{\alpha}_{O_j}(t) =\tr[{(\tau_{O_j}(t))}^{\alpha}] \,,\quad
S^{\alpha}_{O_j}(t) = \frac{1}{1-\alpha}\log(\mathcal{T}_{\alpha}) \label{eq:gen_reny_entro} \,.
\end{equation}
Although these quantities currently lack a clear interpretation within quantum information theory, recent work has shown that the rank of such a reduced transition matrix is related to operator entanglement---a measure of the complexity of simulating the time evolution of local operators in the Heisenberg picture using matrix product states (MPS)~\cite{carignano2024a}.

\section{Generalized temporal entropies as the measurement of local operators after quenches\label{sec:double_quench}}

In this section, we present our first result demonstrating that the generalized temporal R\'enyi entropies
 defined in \cref{eq:gen_reny_entro} can be extracted by measuring the expectation value of Hermitian operators after a double quench experiment on a replicated system.
The operator is the same as the one we have used to ensure that the real-time path integral does not trivialize, as discussed in the construction of generalized entropies. Note that such an operator acts simultaneously on all replicas.

\begin{figure}
\includegraphics[width=\linewidth]{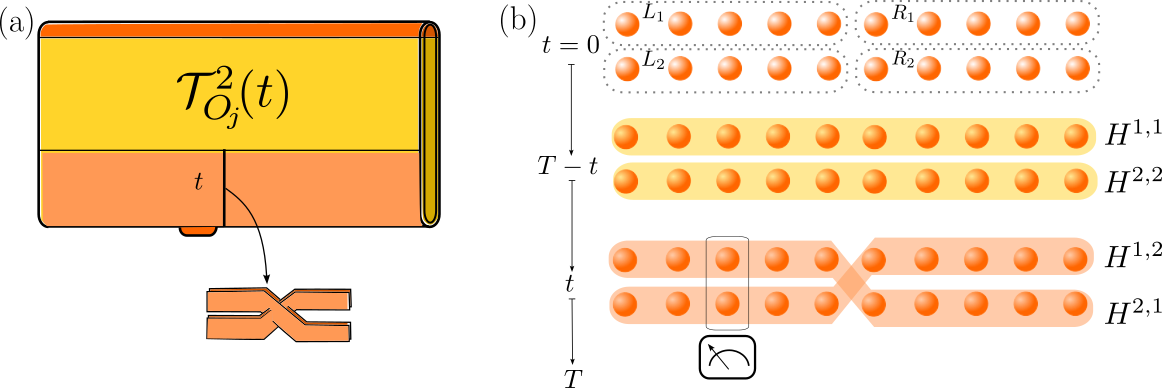}
\caption{\label{fig:setup} (a) The generalized temporal purity $\mathcal{T}_{O_j}^2(t)$ is represented by a path integral on a two-sheeted Riemann surface, where transitions between the sheets occur only along the temporal cut. (b) This path integral can be evaluated experimentally by measuring the expectation value of a local operator acting on a replicated system as described by \cref{eq:quench1} and \cref{eq:quench2}. The two replicas are represented by rows of individual constituents (orange spheres) one for each replica. The initial states of the replicas are quenched twice. First, we apply a Hamiltonian without interactions among the replicas, $H^{1,1}+ H^{2,2}$ (yellow shadows). After time $T-t$, the Hamiltonian is quenched again to $H^{1,2}+H^{2,1}$ (orange shadows), and the replicas interact across the cut. The system evolves for a further time $t$, after which the desired operators are measured simultaneously on both replicas.}
\end{figure}

For simplicity, we explicitly describe how to compute the generalized purity $\mathcal{T}_{O_j}^2(t)$, as higher order Rényi entropies can be obtained analogously. The quantity $\mathcal{T}_{O_j}^2(t)$ is encoded by the contraction of a tensor network wrapped around a two-sheeted Riemann surface, where each sheet encodes one copy of $\tau_{O_j}(t)$.
The multiplication of the two $\tau$ matrices and the trace in \cref{eq:gen_reny_entro} connect the two sheets along the temporal cut in a non-touching crossing, as sketched in \cref{fig:setup}(a).

To identify the  quench protocol for $\mathcal{T}^2$, we define a system consisting of two replicas of the initial state, each split into a right and left part.
This conceptually creates a two-by-two grid where left and right indicate the columns and, $1$ and $2$ indicate the rows of replicas, see \cref{fig:setup}(b). 
We divide the time evolution in two phases:
During the first phase, the two replicas, $1$ and $2$, evolve separately, respectively under Hamiltonians \(H^{1,1} = H^{1}_R + H^{1}_L + H^{1,1}_{L,R}\) and \(H^{2,2} = H^{2}_R + H^{2}_L + H^{2,2}_{L,R}\).
The term \(H_{L,R}^{n,n}\) describes interactions across the left-right cut.
During the second phase of the time evolution, the interaction terms across the left-right cut swap their partner with the other replica.
The entire system thus evolves under the Hamiltonians \(H^{1,2} = H^{1}_R + H^{2}_L + H^{1,2}_{L,R}\) and \(H^{2,1} = H^{2}_R + H^{1}_L + H^{2,1}_{L,R}\), respectively.
Note that the local operator $O_j$ is also replicated and becomes $\mathcal{O}_j\equiv O_j^1\otimes O_j^2$ for the entire system.

To interpret $\mathcal{T}_{O_j}^2(t)$ as a measurement after a quench, we begin with the two replicas in some initial state, and abruptly quench the Hamiltonian of the two replicas to \(H_0 = H^{1,1} + H^{2,2}\), allowing the system to evolve for time \(T-t\). At time \(T-t\), we quench further the Hamiltonian to \(H_{T-t} = H^{1,2} + H^{2,1}\) and allow the system to evolve for the remaining time \(t\). After a total evolution time \(T\), we measure the desired operator \(\mathcal{O}_j\) simultaneously on both replicas, \(\expval*{O_j^1 \otimes O_j^2 (T)}\).
To compute $\mathcal{T}_{O_j}^2$, we also require the correct normalization factor, \(\langle O_j(T) \rangle^2\). This can be obtained through a separate quench on one replica (say, the first), which is quenched to the Hamiltonian \(H^{1,1}\) and allowed to evolve for the full time \(T\). The square of this measurement provides the normalization factor.

The final results thus read,
\begin{equation}
\tr({\tau(t)}_{O_j}^2)\equiv \frac{\tr\left(O^1_j \otimes O_j^2
\rho^{1,2}(T)\right)}{\langle O_j(T)\rangle^2},\label{eq:quench1}
\end{equation}
with
\begin{equation}
\rho^{1,2}(T)\equiv W^{1,2}(T)\left(\rho_0^1\otimes \rho_0^2 \right) W^{1,2}(T)^\dagger, \label{eq:quench2}
\end{equation}
where 
$$W^{1,2}(T)=
\left(U^{1,1}(T-t) \otimes U^{2,2}(T-t)\right) \left( U^{1,2}(t)\otimes U^{2,1}(t)\right) 
$$
is constructed from the corresponding Hamiltonians $H^{1,1}, H^{2,2},H^{1,2}, H^{2,1}$ as defined above.
This approach generalizes for arbitrary $\mathcal{T}^\alpha$, where instead of evolving two replicas, we evolve \(\alpha\) replicas and allow them to interact in pairs across the cut during \(t\).
Therefore, we have just shown how  generalized temporal entropies are equivalent to the measurement of the local operator $\mathcal{O}_j$ after a  double quench experiment on a replicated system.
We now explicitly address the consequences of such equivalence.

\subsection{Generalized temporal entropies are real-valued}

As already mentioned, the transition matrices defined in \cref{eq:red_tm} are complex matrices. Even when they can be diagonalized, their spectrum is generally complex, and so, in principle, are their generalized temporal entropies.
However, identifying generalized temporal entropies with the measurement of a Hermitian operator after a quantum quench implies that these entropies must be real.
This is consistent with results on complex generalized temporal entropies in CFTs~\cite{carignano2024,nakata2021,doi2023a,doi2023b,narayan2023,narayan2023b}. There, the computation is performed at the level of a matrix element $\bra{\psi} U(T)\ket{\psi}$, yielding $S(t,T)_{\text{CFT}} = f(t/T) + i \kappa$, where $\kappa$ is a constant and all the time dependence resides in the real part.
Without loss of generality, we can assume the initial state $\ket{\psi}$ to be a product state. Defining the operator $\mathcal{O} = \ketbra{\psi}{\psi}$, we note that although $\mathcal{O}$ is nonlocal, it does not spoil the $2D$ space-time structure of the tensor network since it remains a product of local operators.
For this choice, the expectation value $\langle \mathcal{O}(T) \rangle$ factorizes as
\begin{equation}
\langle \mathcal{O}(T) \rangle = \bra{\psi} U(T) \ket{\psi} \bra{\psi} U^\dagger(T) \ket{\psi},
\end{equation}
and thus the generalized temporal entropy becomes a sum of two contributions: one from $\bra{\psi} U(T) \ket{\psi}$ and one from $\bra{\psi} U^\dagger(T) \ket{\psi}$. These two pieces differ only by a sign in front of $t$.
As a result, we can apply the analytic continuation approach for each piece independently: for the first, we take $T \to i\beta$, and for the second, $T \to -i\beta$. Summing the two contributions gives
\begin{equation}
S = f(t/T) + i \kappa + f(t/T) - i \kappa = 2\, \text{Re}\, S(t,T)_{\text{CFT}}.
\end{equation}
This confirms our general claim: the generalized temporal entropies after a quantum quench discussed in this work are real-valued.

\subsection{Generalized temporal entropies as dynamical probes of geometric quenches}\label{sect:pump-probe}

The second implication of our construction is that generalized temporal entropies encode the system's response to a geometric change in a replicated system.
To frame this idea, it is useful to draw an analogy with standard pump-probe experiments, where a system is first excited (pumped) via a perturbation—such as by shining light or neutrons onto it—and later probed by measuring specific operators. Such techniques are at the heart of spectroscopy in two-dimensional materials.
Our results imply that generalized temporal entropies can be interpreted similarly. In our case, the pump phase corresponds to the extended period during which the two replicas are coupled together, exciting the system by populating specific modes of the replicated structure. It is only by measuring local operators at later times that we can probe the effect of this initial excitation and analyze the system's dynamical response, Fig.~\ref{fig:quench_fig}(b).
Thus, the local operators we consider act as probes of the geometric perturbation, and by moving them in space and tuning the time duration of the first phase of the quench, we obtain a complete space-time picture of the replicated system’s response to the quench, Fig.~\ref{fig:quench_fig}(c). As in traditional pump-probe experiments, different operators are sensitive to different excitation branches: for example, charge-neutral operators cannot probe charged currents, while charged operators may be blind to spin or energy transport.
In principle, a full characterization of the response to the initial pump requires measuring all possible operators.
In practice, however, the symmetries of the problem guide the choice, and a few well-chosen operators are typically sufficient to capture the relevant physical effects.
These ideas are fully developed in the following section which aims at using geometric quenches as a probe to discern integrable system from non-integrable ones.

\subsection{The double quench as way to measure generalized temporal entropies in experiments}
\begin{figure}
  \includegraphics[width=\linewidth]{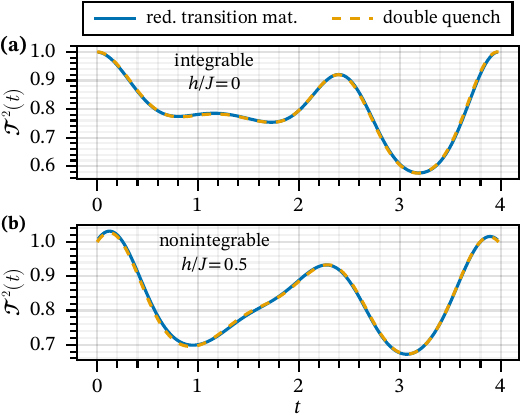}
  \caption{\label{fig:num_check_quench}%
    Temporal purities $\mathcal{T}^2_{\ketbra{0}{0}}(t) = \tr(\tau_{\ketbra{0}{0}}( t ))^2$ for the quench dynamics starting in a product state and evolved up to \(TJ=4\). Time in the x-axis is measured in units of inverse energy (\(1/J\)). Note that \(O = \ketbra{0}{0}\) is evaluated in the middle of the chain for the finite size systems (dashed yellow lines).
    Results shown are time evolved with the Hamiltonian \eqref{eq:ham} with \(g/J=0.4\) and \textbf{(a)} \(h/J=0\) (integrable) and \textbf{(b)} \(h/J=0.5\) (non-integrable).
    We compute $\mathcal{T}^2(t)$ for all possible temporal bi-partitions \(0 \leq t \leq T\) yielding reduced transition matrices \(\tau_{\ketbra{0}{0}}( t )\) and plot $\mathcal{T}^2$ as a function of \(t\).
    The direct calculation of $\mathcal{T}^2(t)$ via the direct tensor network contraction of the path-integral in the thermodynamic limit are plotted using the blue solid lines, while the results of simulating the experimental quench protocol \cref{eq:quench1} for a system made by \(N=40\) spin by dashed yellow lines.
    As expected the second Rényi entropy $\left(S^2_{\ketbra{0}{0}}(t) = -\log(\mathcal{T}^2_{\ketbra{0}{0}}(t))\right)$ vanishes identically at the beginning \(t=0\) and at the end of the evolution \(t=T\), but it oscillates for bi-partitions in between. There is perfect agreement between the two calculations.
    We observe near perfect agreement between the two methods that just boil down to computing the same object using two different techniques.
  }%
\end{figure}
A consequence of our identification of generalized temporal entropies with the measurement of local-operators after an appropriately designed quench on a replicated system is that
such  a measuring process is feasible on most state-of-the-art quantum simulators, including platforms such as trapped ions or cold atoms, as we will discuss in Sec.~\ref{sec:exp}.
We verify this protocol numerically by simulating the Ising model with both transverse and parallel fields, described by the Hamiltonian:
\begin{equation}
 H=- J \sum_{i=1}^{N-1} \sigma_{x,i}\sigma_{x,{i+1}} +\sum_{i=1}^N \left(g\sigma_{z,i} + h \sigma_{x,i}\right), \label{eq:ham}
\end{equation}
which is exactly solvable for $h=0$. On this line, the system is described by free Majorana fermions and as such is integrable. The model has critical points at $g=\pm J, h=0$ in the 2D Ising universality class, while the rest of the phase diagram is gapped, with a line of first order transitions terminating at the critical points~\cite{pfeuty1969,fogedby1978,ovchinnikov2003}.
In \cref{fig:num_check_quench}, we compare numerical simulations of the quench with direct calculations of the $\mathcal{T}^2$ in two different points of the phase diagram, one along the integrable line in the ferromagnetic phase, and the other in the non-integrable disordered phase.
The dotted lines show the results of matrix product states (MPS) simulations of the double quench experiment.
The operator $\mathcal{O}_{N/2}= \tfrac{(1+\sigma_z)}{2}$, located at the temporal cut, is measured simultaneously on both replicas at the end of the quench.
Each replica contains $N=40$ spins and the total evolution time is $T=4J$. The time evolution of the system is simulated through time-evolving block decimation (TEBD)~\cite{vidal2003,schollwock2011}. By partitioning the time into $T-t$ and $t$ with $0\le t\le T$, and applying the double quench experiment for each $t$, we obtain $\mathcal{T}^2$ for all $t\le T$.
Alternatively, we can contract the corresponding tensor network on the two sheets replicated Riemann surface directly in the thermodynamic limit, following the algorithms described in ~\cite{carignano2024a}. The results are presented as solid lines in the same plot. The perfect agreement between the two lines demonstrate that the two approaches compute the same mathematical object.

\subsection{Continuum time limit of generalized temporal entropies}

We now emphasize an important consequence of the correspondence established in ~\cref{eq:quench1}, which maps the computation of generalized temporal purity $\mathcal{T}_{O_j}^2(t)$ to the measurement of a local operator after a quench: this mapping implies that the generalized temporal entropy does not suffer from ultraviolet (UV) divergences when the temporal lattice spacing $\delta t$ is sent to zero.

It is well-known that the expectation value of a local operator under continuous-time evolution can be accurately approximated by a Trotter discretization, and that such approximations can be systematically improved. This ensures that taking the continuum time limit does not introduce divergences in the observable.
For the second R\'enyi entropy to diverge, the generalized purity in \cref{eq:quench1} would have to vanish. This can occur under two possible conditions:
(i) the numerator vanishes while the denominator remains finite, or
(ii) the denominator diverges while the numerator remains finite or vanishes.
Case (ii) can be excluded, since the denominator is the expectation value of a bounded local operator and thus cannot diverge. Case (i) can occur—for example, the expectation value of a local operator may vanish exactly at finite $\delta t$. However, this does not signal a UV divergence.
A true UV divergence would require that the generalized purity decreases to zero \emph{only} in the limit $\delta t \to 0$, even when it is finite at any nonzero $\delta t$. This is incompatible with the known behavior of Trotterized time evolution: for example, a second-order Trotter approximation yields errors of order $\mathcal{O}(\delta t^2)$, so if the expectation value is finite at finite $\delta t$, it must remain finite as $\delta t \to 0$.
This property of generalized temporal entropies ensures that our lattice-based calculations can be meaningfully compared with experimental results from analog quantum simulators, where time is inherently continuous.

It is important to note that our conclusions rely crucially on the fact that we are working with spatial lattices. If we were instead to take the continuum limit in space as well, the arguments concerning local operators would need to be revisited. In that case, the expectation value in the denominator of \cref{eq:quench1} could, in principle, diverge, potentially leading to a genuine UV divergence.

\subsection{Entropic meaning of generalized temporal entropies}  \label{sec:entropic}

The quantities we have discussed so far do not necessarily possess a strict interpretation as entropies. From a field-theory point of view, we are computing the correlation function of a twist field (the construction used to obtain the R\'enyi-two entropy, inserted at the end of the temporal interval \cite{doyon2025}) with an operator $\mathcal{O}$ that prevents the collapse of the full tensor network and, as we have just seen, probes the effect of the twist field on the subsequent dynamics. Field-theoretical entropic quantities on the other hand are in general function of twist fields alone.

\begin{figure}[t]
    \centering
    \includegraphics[width=\linewidth]{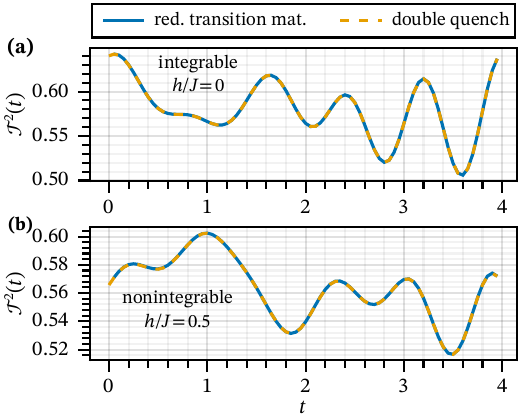}
    \caption{\label{fig:purity_after_the_quench}
        Mixed generalized spatio-temporal purity
        $\mathcal{T}^2_{S_{12}(r)}(t) = \mathrm{tr}[\tau_{S_{12}}(t)]^2$
        for the quench dynamics starting from a product state and evolved up to $TJ = 4$. Time in the x-axis is measured in units of inverse energy (\(1/J\)).
        The swap operator $S_{12}$ between the two replicas is applied to the spins in the middle of each chain.
        The dynamics is generated by the Hamiltonian~\eqref{eq:ham} with $g/J = 0.4$ for
        \textbf{(a)} $h/J = 0$ (integrable) and
        \textbf{(b)} $h/J = 0.5$ (non-integrable).
        For each temporal bipartition $0 \le t \le T$, we obtain reduced transition matrices $\tau_{S_{12}}(t)$ and plot $\mathcal{T}^2(t)$ as a function of time.
        The behavior closely mirrors that observed for local operators, with the important distinction that $\mathcal{T}^2(t)$ is now guaranteed to remain below one for all scenarios.
    }
\end{figure}

By noticing that \cref{eq:quench1} involves a ratio, one immediately understands that the resulting quantity can in principle exceed one, even though both the numerator and the denominator can be considered positive (as they involve positive-definite operators) and smaller or equal to one under an appropriate normalization. Indeed, whenever the expectation value of the square of the operator on a single copy is smaller than the expectation value of the tensor product of the operator acting on two copies after the quench, the generalized purity exceeds one. Such a regime is visible, for example, in panel (b) of Fig.~\ref{fig:purity_after_the_quench}, close to the beginning and the end of the evolution. In this situation, the quantity no longer admits a direct interpretation as entropy.

This is not an issue in itself, after all, we are dealing with generalized entropies. Still, we note that we can easily obtain a genuine entropy in our setup by replacing the tensor product of the local operator $\mathcal{O}$ on the two copies with a swap operator. In the field-theory sense, we substitute the local operator for an extra (pair of) twist field, so this quantity should now have a purely entropic interpretation.

In practice, rather than using a local operator, we are now probing the quench encoding the generalized temporal entropy via its effect on the purity of a single site. We thus measure the mixed spatio-temporal generalized purity of a subsystem consisting of the reduced density matrix of one spin (the one on which the original operator acted) together with the reduced transition matrix of the temporal degrees of freedom.

Following the same procedure as before, we need to normalize such mixed reduced density-transition matrix by dividing by its trace. The latter is just the full contraction of the tensor network, which now evaluates always to one. We thus completely solve the problem of small denominators in the expression of the generalized temporal entropies \cref{eq:quench1} which lead to values of the generalized entropy larger than one.

Furthermore, the purity of any reduced density matrix after the quench is strictly smaller than one, which guarantees that the resulting quantity has a bona fide entropic meaning. This behavior is illustrated in Fig.~\ref{fig:purity_after_the_quench}.

Such a construction opens also new avenues of research, in which one can try to characterize the purity of blocks of different sizes after the quenches encoding the generalized entropies.

\section{Footprints of integrability in temporal entropies}\label{sec:footprint-integrability}

\begin{figure*}
  \includegraphics[width=\linewidth]{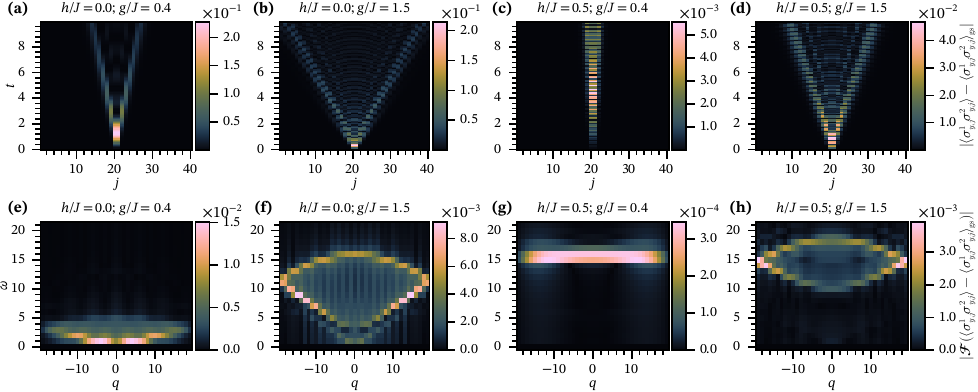}\\
 \caption{\label{fig:fourier_ana_1}%
   Temporal purities $\mathcal{T}_{O_j}^2(t)$ with \(O_j = \sigma^y_j\) for systems made up of \(N=40\) spins starting in the ground state of the Ising model \cref{eq:ham}. Panels (a)--(d) show their space time profile (with time measured in units of inverse energy (\(1/J\))), while the corresponding Fourier transforms obtained via Eq.~\eqref{eq:fourier} are displayed in (e)--(h), with frequency measured in units of energy (\(J\)). Integrable cases: (a) and (e) lie in the ferromagnetic phase, \(g/J=0.4\) and \(h/J=0\);  (b) and (f) in the paramagnetic phase  \(g/J=1.5\) and \(h/J=0\).  Non-integrable cases: (c) and (g) show data for \(g/J=0.4\) and \(h/J=0.5\); (d) and (h) correspond to \(g/J=1.5\) and \(h/J=0.5\). The bright regions in (e) and (f) close to $\omega=0$ and $q=0$ unveil the appearance of soft modes in the integrable cases which are absent in (g) and (h) corresponding to the non-integrable cases.
  }
\end{figure*}

In this section, we provide the first application of the measurement of generalized temporal entanglement, by observing that 
 the momentum-frequency properties of $\mathcal{T}^2$ vary between integrable and non-integrable dynamics.
We simulate the quench leading to $\mathcal{T}^2$ described in Sec.~\ref{sec:double_quench} using tensor-network techniques.
We consider a chain made up of $N=40$ spins and characterize $\mathcal{T}^2$ for the out-of-equilibrium dynamics of the ground state of the Ising model Hamiltonian of \cref{eq:ham} in the gapped regions of the phase diagram.
We compute $\mathcal{T}^2$ of $\tau_{O_j}(t)$ obtained from a cut located at the center of the chain.
As discussed in \cref{sect:pump-probe}, we can vary the location of the operator ${O_j}$ to probe the spatial response of the system to the perturbation induced by coupling the different replicas during the second phase of the quench protocol.
To simplify the setup, we choose as an initial state the ground state of the Hamiltonian
of the separated replicas.
Such ground state of $H$ in \cref{eq:ham} can be obtained through standard  density matrix renormalization group (DMRG)~\cite{white1992,schollwock2011}.
During the first part of the quench the copies evolve separately, and as a result nothing happens.
The evolution starts only in the second phase of the quench, when we exchange the left and right part of the replicas. As a result, we only have to keep track of one of the two time scales, $t$.
For every $t$ we measure the local operator along the chains and this gives rise to a space-time picture of $\mathcal{T}^2$. At this stage we can  apply the standard quench-spectroscopy~\cite{kormos2017,lagnese2021,villa2019,villa2020,chanda2024}. The protocol unveils the basic fact that the excess energy generated by the quench is redistributed among the low energy excitations of the system.
These excitations can be identified by performing a frequency momentum analysis of the space-time profile of $\mathcal{T}^2$, which gives their dispersion relations.

The space-time results for the $\mathcal{T}^2_{{\sigma_x}_j}(t)-1$ are presented in panels (a), (b), (c), and (d) of Fig.~\ref{fig:fourier_ana_1} for times up to $TJ=9$. As expected they have the space-time shape of what we measure after a local quench in the center of the system, where the two half systems are exchanged.
Importantly, different operators measure different aspects of the space time quench, but their frequency-momentum transformation provides qualitatively similar pictures.
The system we are analyzing does not contain any conserved quantities other than the energy, thus all operators couple similarly to the same low-energy modes.
By observing our numerical results, it is clear that the integrable system $g/J =0.4$,  $h/J=0$ and $g/J =1.5$,  $h/J=0$ behave markedly differently from the non-integrable ones $g/J=0.4$, $h/J=0.5$ and $g/J=1.5$, $h/J=0.5$.
We perform the frequency--momentum analysis which allows distinguishing this difference qualitatively in terms of the excitation structure of the two systems by providing access to the low energy dispersion relation of the excitations of the replicated system~\cite{kormos2017,lagnese2021,villa2019,villa2020}.

We define
 \begin{equation}
 \mathcal{F}_{\tau^2}(q, \omega) = \frac{2 \pi}{L T} \delta t \sum_{j=1}^{N} e^{-i q (j-\frac{N}{2})} \sum_{t=0}^{T} e^{-i \omega t} \left(\mathcal{T}^2_{O_j}(t) - 1\right).
 \label{eq:fourier}
\end{equation}
Panels~(e), (f), (g) and (h) of Fig.~\ref{fig:fourier_ana_1} show contour plots of Eq.~\eqref{eq:fourier} with the zero-frequency and momentum data omitted, as these would be unreliable due to the finite space-time extent of the simulations.
Panels~(e) and (f) reveal the appearance of a soft mode (the bright line in the lower part), 
which is absent in the dispersion relation of the individual copies due to the gapped nature of the system.
This soft mode is a clear footprint of the replicated system.
Conversely, panels~(g) and (h) in the non-integrable regime shows the absence of any soft mode.

To demonstrate that the soft mode is specific to the gapped integrable region of the phase diagram, we focus on the frequency component of Eq.~\eqref{eq:fourier},
\begin{equation}
 \lambda(\omega) = \frac{2 \pi}{T} \delta t \sum_{t=0}^{T} e^{-i \omega t} \left({ \mathcal{T}^2_{O_{N/2}}(t) - 1}\right)
 \label{eq:lambda}.
\end{equation}

\begin{figure}
 \includegraphics[width=\linewidth]{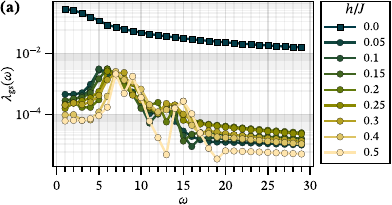}\\
 \includegraphics[width=\linewidth]{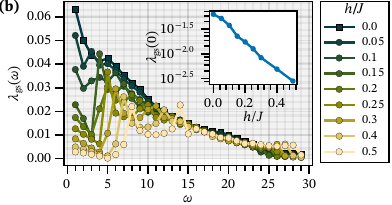}\\
 \includegraphics[width=\linewidth]{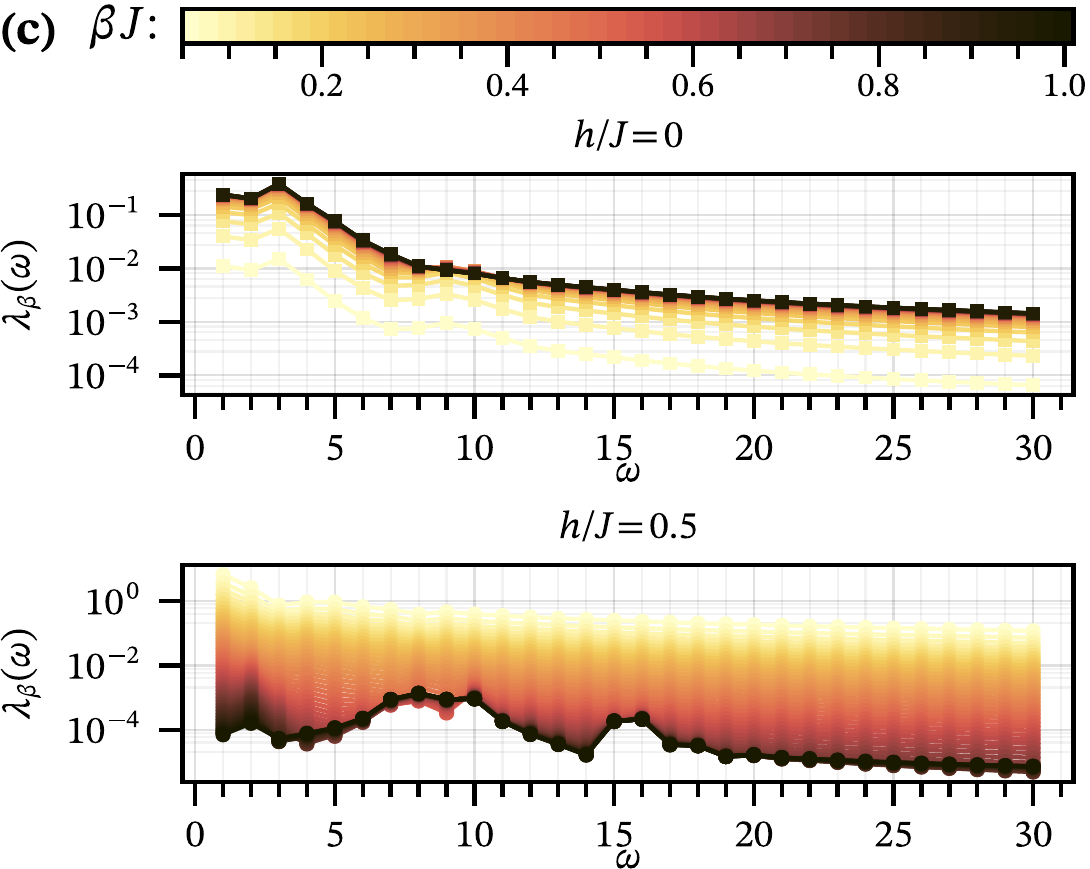}
 \caption{\label{fig:fourier_ana_2}%
  We show in panel (a) the frequency part of the response $\lambda(\omega)$ (with $\sigma_x$ correlators instead of $\sigma_y$) defined in Eq.~\eqref{eq:lambda} for several values of the symmetry breaking field $h/J$. Frequency in the x-axis of the panels is measured in units of energy (\(J\)).
  We see that in the ferromagnetic phase, for $g/J=0.4$, the soft mode disappears as soon as one moves away from the integrable case \(h/J=0\).
  In panel (b) We show the frequency part of the response (with $\sigma_z$ correlators instead of $\sigma_y$) for \(g/J = 1.5\), in the paramagnetic phase, for several values of $h/J$. In this case the soft mode does not disappear, but it is exponentially suppressed as \(h/J\) increases (inset).
  Similarly, for quench dynamics initialized with a thermal state, we show in panel (c) the same quantity $\lambda(\omega)$
  where one can observe that the results rapidly tend towards their ground state counterpart for already moderate values of $\beta J \simeq 1$. This means that these results should be easy to observe in available experiments based on quantum simulators.
 }
\end{figure}
We start characterizing the departure from the ferromagnetic phase due to $h$ in panel (a) of Fig.~\ref{fig:fourier_ana_2}. There we plot $\lambda(\omega)$ for $g/J=0.4 $ and various choices of $h/J$ encoded by different colors to track the behavior of the soft mode as integrability is broken. The soft mode is identified by a non-vanishing value of $\lambda(\omega)$ close to $\omega=0$. As $h/J$ is increased from zero, the soft mode quickly disappears.
For finite $h/J$, the curves cluster towards a peak between $5 \leq \omega \leq 10$, signalling the absence of the soft mode.
Our finite-precision data are consistent with $\lambda(\omega)$ vanishing as $\omega \to 0$ for all $h/J \neq 0$.
While one could attribute this fact to the strong confinement of mesons  at the specific non-integrable point we are considering of the Ising model, we have repeated the same analysis
for  $g/J =1.5$, where the confinement of mesons does not influence the dynamics after the quench ~\cite{kormos2017}.
In panel (b) of Fig.~\ref{fig:fourier_ana_2}, we perform the same analysis in the disordered phase at $g/J = 1.5$. The situation here is more complex, as the soft mode persists in the non-integrable regime for small values of $h/J$.
However, as shown in the inset, its strength decreases exponentially as a function of $h/J$ to zero. As a result, for $h/J=0.15$ representing $10\%$ of the total interaction, the zero mode is significantly suppressed with respect to a finite mode that appears at around $\omega=5$. Thus, the weak zero mode can be considered a signature of proximity to integrability rather than a feature of non-integrable dynamics.

Our results also remain robust at finite temperature, as depicted in panel (c) of Fig.~\ref{fig:fourier_ana_2}. Here, we repeat the analysis performed for Fig.~\ref{fig:fourier_ana_1}, now in the thermodynamic limit, starting from a thermal state and at finite temperature. Different lines encode various values of the inverse temperature $\beta$, and we compute $\lambda(\omega)$ using Eq.~\eqref{eq:lambda}. We observe that $\lambda(\omega)$ rapidly converges to the ground state behavior as $\beta$ increases.
For modest value of $\beta$, specifically $\beta J \simeq 1$, the soft mode stabilizes in the integrable case, while finite-frequency peaks become more prominent in the non-integrable case.
We attribute the stability of these results at finite temperature to the presence of a gap in the systems before coupling. This robustness suggests that our findings should be observable in current quantum simulation experiments.
As a result, we can claim that by breaking integrability, the soft mode is exponentially suppressed with respect to $h/J$ in all the cases we have analyzed.
An interpretation of the appearance of a soft mode for integrable systems can be given in terms of fragility of integrability to the change of geometry and, in that case, it would imply that what we are observing is generic and not restricted to the Ising model studied here.
As a consequence of this change in geometry, the geometric quench which encodes the generalized purity,
immediately breaks integrability, leading to a strong response, and the appearance of a soft mode associated with it.
In order to notice such change of geometry, the correlation length of the initial state should be sufficiently small. This implies that we need to start from a gapped initial ground state. In critical systems, universality takes over, and we would be unable to distinguish a two rungs ladder from  a one dimensional chain.
The same interpretation allows understanding the absence, or exponential suppression, of the soft mode in non-integrable systems. These are not specific to one dimension, and given that we are characterizing gapped systems, we expect that the local nature of the quench does not inject enough energy to change drastically the dispersion relation of the replicated system with respect to the one of two separated replicas, the system being protected by its energy gap. As a result the response to these quenches is extremely suppressed.

\section{Measuring temporal entropies in experiments}
\label{sec:exp}

\subsection{Strategy for experimentally measuring the purity of a state}
The entanglement entropy of spatial bipartitions is a central diagnostic of many-body quantum systems, and significant effort has been devoted to providing experimental access to it.  The most direct method is to perform full state tomography of the reduced density matrix, which is feasible only for small systems due to its exponential cost in system size. Alternatives include randomized measurements and classical shadows.

A third, highly successful strategy is to directly measure Rényi entropies using multiple copies of the quantum state, in close analogy with what we propose in this work. The first experimental measurement of the purity was performed in a cold-atom setup in Ref.~\cite{islam2015,kaufman2016}. The key idea is the identity
\begin{equation}
\textrm{tr}(\rho_A^2) =\bra{\psi_1\otimes\psi_2} S^A_{1,2}\ket{\psi_1\otimes\psi_2} \,,
\end{equation}
 where $S^A_{1,2}$ is the swap operator acting only on subsystem $A$ of two identical copies $\ket{\psi_1}$ and $\ket{\psi_2}$ of the full system.

The experimental challenge therefore reduces to preparing two identical replicas and implementing a swap operation on a chosen region.
 In Ref.~\cite{islam2015,kaufman2016}, this was achieved using single-site addressing and by coherently mapping symmetric and antisymmetric combinations of bosonic operators
 $a_s =\frac{1}{2}\left(a_1+a_2\right)$, $a_a =\frac{1}{2}\left(a_1-a_2\right)$, onto physical modes of a two-leg ladder via a controlled beam-splitter operation (inter-copy tunneling for a calibrated time). The occupation-parity of these modes directly yields the eigenvalue $\pm1$
of the local swap operator; the eigenvalue of the full swap is the product of the local contributions.

This experiment demonstrates unambiguously that the type of swap operation required in our proposal can be implemented while maintaining full coherence of the quantum state.

\subsection{Experimental setups and requirements}
In close analogy with spatial entanglement measurements, our quench-based protocol for temporal entropies requires only standard capabilities that are already available in several quantum-simulation platforms:
 (1) the ability to prepare two identical replicas of the system, (2) local control over the coupling between replica to perform the partial swap, and (3) the possibility to  measure  either local observables or momentum-resolved quantities.

Notice that, in contrast to standard protocols for measuring spatial Rényi purities, our scheme for generalized (temporal) purities does not necessarily require measuring the swap operator itself.

Instead, one only needs to implement the swap and then perform the much simpler task of measuring local or momentum-resolved observables, as discussed in the previous sections. One can still decide to measure also the swap operator, if interested in a quantity with a clear interpretation as entropy as discussed in Sec. \ref{sec:entropic}. In the latter case, one still only needs to measure the swap acting on a single constituent of each replica.

These ingredients are currently accessible in ultra-cold atoms in optical lattices, Rydberg arrays, trapped ions, and superconducting qubit platforms~\cite{gross2017,daley2022,bernien2017,monroe2021,rosenberg2024,wienand2024}.

For example, in cold-atom setups, a promising realization involves a double-well potential~\cite{wienand2024}, where each well encodes one replica. Real-time control over the tunneling between wells enables a switch from independent to interacting dynamics, thereby implementing the central quench of our protocol. Site-resolved control and detection allow both local measurements and Fourier transforms.

An alternative setup is the one used to study many-body scarring in Ref.~\cite{su2023}, where a spin-dependent optical super lattice is engineered in a two-dimensional configuration with one short (two-site) direction and one long direction. The short direction naturally encodes the replicas. The same beam-splitter technique used there to measure the time-evolving purity can implement our protocol, transferring atoms between replicas in a controlled fashion. Furthermore, with tilted lattices and metastable initial states~\cite{sachdev2002}, this setup can simulate Ising-like dynamics compatible with our scheme.

Another cold-atom realization uses double wells with time-controlled tunneling. One could initialize two disconnected replicas by raising the potential barrier between wells, then freeze the system by raising the global potential. By briefly lowering the inter-well barrier on half of the chain, atoms can be transferred coherently to the other well. If the ramp is fast compared to decoherence times, the system preserves phase coherence. The replicas can then be re-separated and dynamics resumed, now with partial inter-replica interaction as required by our protocol.
This approach is platform-agnostic and adaptable to Rydberg atom arrays and trapped-ion systems. For instance, in Rydberg or tweezer setups, individual atom rearrangement and pairwise control can reproduce the required, this experimental setup is illustrated in~\cref{fig:quench_fig}(a) and (b).

Yet another approach encodes replicas in two internal states (species) of the same atom, labeled $a$ and $b$. Starting with species-specific optical potentials and $U_{ab} \approx 0$ via Feshbach tuning, one can initialize disconnected dynamics.
The inter-species coupling can then be activated via spin-changing tunneling at a target site using Raman processes---techniques already developed for simulating gauge theories~\cite{schweizer2019,yang2020}, this experimental setup is illustrated in~\cref{fig:setup}(b).

Finally, continuum implementations may also be possible using coherent wave packet techniques demonstrated in Ref.~\cite{viermann2022}.
Trapped ions also provide natural flexibility, with replicas encoded either in spatial modes or internal states. Spin-spin interactions mediated via collective vibrational modes allow precise control over connectivity~\cite{monroe2021,guo2024b,qiao2024}. Similarly, Rydberg arrays~\cite{barredo2016,labuhn2016,scholl2021a} and programmable superconducting circuits~\cite{semeghini2021,rosenberg2024} offer high tunability and readout access, making them strong candidates for implementing our protocol.

\section{Conclusions and Outlook}
\label{sec:conclusions}
In this work, we
have proposed a quench protocol that provides experimental access to generalized temporal entropies, allowing to extract information on temporal correlations of a system similarly to a pump-probe setup. This protocol requires at least two replicas when measuring the generalized temporal purity $\mathcal{T}_{O_j}^2(t)$ or the second Rényi entropy. 

Besides opening a new experimental avenue by enabling, for the first time, the experimental characterization of generalized temporal entropies, we also 
unveiled several properties of these quantities, such as that they are real and finite for systems defined on a lattice evolving in continuous time, 
and discussed their interpretation as probe of the systems to geometric quenches.
These results pave the way for the experimental characterization of temporal entanglement, whose importance spans multiple areas of theoretical physics, not only as a critical measure of complexity in the classical simulation of out-of-equilibrium time evolution, but also as a probe providing direct access to bulk geometric properties in holographic field theories and universal information about critical systems in Loschmidt echo quenches.

We observed that generalized temporal entropies are sensitive to the dynamical properties of the models considered: in particular, in the  gapped integrable regime of the Ising model, we observe the emergence of a soft mode in response to the quench.
This soft mode could be a signature of the fragility of integrability, to changes in geometry induced by the double quench we need to implement to measure the generalized entropies. Given that integrability is primarily restricted to one-dimensional systems, defining the dynamics on the replicated Riemann surface,  which starts to explore an extra dimension, immediately breaks it. The soft mode may be a manifestation of such breaking.  This interpretation suggests that our results can apply beyond the Ising model analyzed here, a possibility we are currently investigating.

If this is the case, our findings  offer a potential early diagnostic for distinguishing integrable from non-integrable dynamics, addressing an unresolved experimental challenge. While integrable and non-integrable systems are known to evolve into different equilibrium states, observing these differences typically requires waiting for the system to equilibrate. Such equilibration can occur at late times, often beyond the coherence time of current experimental platforms.

Many intriguing questions remain open, such as the relationship between generalized temporal entanglement and spatial entanglement, or its connection with other measures of temporal entanglement and witnesses studied in the foundations of quantum mechanics. It would also be interesting to explore whether there are extensions of the standard Bell notion of non-locality that necessitate the presence of non-vanishing generalized temporal entanglement.

During the Refereeing process of our manuscript, a new pre-print was announced~\cite{milekhin2025a}, which proposes a closely related concept of entanglement in time and a quantum circuit allowing its measurement on a quantum computer.
In specific cases, the two constructions coincide, and it would be interesting to establish a precise connection between them in the future.

\section{Acknowledgments}
We acknowledge support from the Proyecto Sinérgico CAM Y2020/TCS-6545 NanoQuCo-CM, the CSIC Research Platform on Quantum Technologies PTI-001, and from the Grant TED2021-130552B-C22 funded by MCIN/AEI/10.13039/501100011033 and by the ``European Union NextGenerationEU/PRTR'', and Grant PID2021-127968NB-I00 funded by MCIN/AEI/10.13039/501100011033.

ABC is supported by Grant CQS2301001 from the Project Quantum ENIA. The funding for this grant comes from the Plan de Recuperación, Transformación y Resiliencia en el marco de la Agenda España Digital 2025, from the Ministerio de Asuntos Económicos y Transformación Digital. ABC is also supported by Grant MMT24IFF-01 and SC by Grant C005/24-ED CV1. The funding for these actions/grants and contracts comes from the European Union’s Recovery and Resilience Facility-Next Generation, in the framework of the General Invitation of the Spanish Government’s public business entity Red.es to participate in talent attraction and retention programs within Investment 4 of Component 19 of the Recovery, Transformation, and Resilience Plan.

\bibliography{exp_te.bib}

\begin{thebibliography}{77}%
\makeatletter
\providecommand \@ifxundefined [1]{%
 \@ifx{#1\undefined}
}%
\providecommand \@ifnum [1]{%
 \ifnum #1\expandafter \@firstoftwo
 \else \expandafter \@secondoftwo
 \fi
}%
\providecommand \@ifx [1]{%
 \ifx #1\expandafter \@firstoftwo
 \else \expandafter \@secondoftwo
 \fi
}%
\providecommand \natexlab [1]{#1}%
\providecommand \enquote  [1]{``#1''}%
\providecommand \bibnamefont  [1]{#1}%
\providecommand \bibfnamefont [1]{#1}%
\providecommand \citenamefont [1]{#1}%
\providecommand \href@noop [0]{\@secondoftwo}%
\providecommand \href [0]{\begingroup \@sanitize@url \@href}%
\providecommand \@href[1]{\@@startlink{#1}\@@href}%
\providecommand \@@href[1]{\endgroup#1\@@endlink}%
\providecommand \@sanitize@url [0]{\catcode `\\12\catcode `\$12\catcode `\&12\catcode `\#12\catcode `\^12\catcode `\_12\catcode `\%12\relax}%
\providecommand \@@startlink[1]{}%
\providecommand \@@endlink[0]{}%
\providecommand \url  [0]{\begingroup\@sanitize@url \@url }%
\providecommand \@url [1]{\endgroup\@href {#1}{\urlprefix }}%
\providecommand \urlprefix  [0]{URL }%
\providecommand \Eprint [0]{\href }%
\providecommand \doibase [0]{https://doi.org/}%
\providecommand \selectlanguage [0]{\@gobble}%
\providecommand \bibinfo  [0]{\@secondoftwo}%
\providecommand \bibfield  [0]{\@secondoftwo}%
\providecommand \translation [1]{[#1]}%
\providecommand \BibitemOpen [0]{}%
\providecommand \bibitemStop [0]{}%
\providecommand \bibitemNoStop [0]{.\EOS\space}%
\providecommand \EOS [0]{\spacefactor3000\relax}%
\providecommand \BibitemShut  [1]{\csname bibitem#1\endcsname}%
\let\auto@bib@innerbib\@empty
\bibitem [{\citenamefont {Amico}\ \emph {et~al.}(2008)\citenamefont {Amico}, \citenamefont {Fazio}, \citenamefont {Osterloh},\ and\ \citenamefont {Vedral}}]{amico2008}%
  \BibitemOpen
  \bibfield  {author} {\bibinfo {author} {\bibfnamefont {L.}~\bibnamefont {Amico}}, \bibinfo {author} {\bibfnamefont {R.}~\bibnamefont {Fazio}}, \bibinfo {author} {\bibfnamefont {A.}~\bibnamefont {Osterloh}},\ and\ \bibinfo {author} {\bibfnamefont {V.}~\bibnamefont {Vedral}},\ }\bibfield  {title} {\bibinfo {title} {Entanglement in many-body systems},\ }\href {https://doi.org/10.1103/RevModPhys.80.517} {\bibfield  {journal} {\bibinfo  {journal} {Reviews of Modern Physics}\ }\textbf {\bibinfo {volume} {80}},\ \bibinfo {pages} {517} (\bibinfo {year} {2008})}\BibitemShut {NoStop}%
\bibitem [{\citenamefont {Laflorencie}(2016)}]{laflorencie2016}%
  \BibitemOpen
  \bibfield  {author} {\bibinfo {author} {\bibfnamefont {N.}~\bibnamefont {Laflorencie}},\ }\bibfield  {title} {\bibinfo {title} {Quantum entanglement in condensed matter systems},\ }\href {https://doi.org/10.1016/j.physrep.2016.06.008} {\bibfield  {journal} {\bibinfo  {journal} {Physics Reports}\ }\bibinfo {series} {Quantum Entanglement in Condensed Matter Systems},\ \textbf {\bibinfo {volume} {646}},\ \bibinfo {pages} {1} (\bibinfo {year} {2016})}\BibitemShut {NoStop}%
\bibitem [{\citenamefont {Zeng}\ \emph {et~al.}(2018)\citenamefont {Zeng}, \citenamefont {Chen}, \citenamefont {Zhou},\ and\ \citenamefont {Wen}}]{zeng2018}%
  \BibitemOpen
  \bibfield  {author} {\bibinfo {author} {\bibfnamefont {B.}~\bibnamefont {Zeng}}, \bibinfo {author} {\bibfnamefont {X.}~\bibnamefont {Chen}}, \bibinfo {author} {\bibfnamefont {D.-L.}\ \bibnamefont {Zhou}},\ and\ \bibinfo {author} {\bibfnamefont {X.-G.}\ \bibnamefont {Wen}},\ }\href {https://doi.org/10.48550/arXiv.1508.02595} {\bibinfo {title} {Quantum {{Information Meets Quantum Matter}} -- {{From Quantum Entanglement}} to {{Topological Phase}} in {{Many-Body Systems}}}} (\bibinfo {year} {2018}),\ \Eprint {https://arxiv.org/abs/1508.02595} {arXiv:1508.02595 [cond-mat, physics:quant-ph]} \BibitemShut {NoStop}%
\bibitem [{\citenamefont {Callan}\ and\ \citenamefont {Wilczek}(1994)}]{callan1994}%
  \BibitemOpen
  \bibfield  {author} {\bibinfo {author} {\bibfnamefont {C.}~\bibnamefont {Callan}}\ and\ \bibinfo {author} {\bibfnamefont {F.}~\bibnamefont {Wilczek}},\ }\bibfield  {title} {\bibinfo {title} {On {{Geometric Entropy}}},\ }\bibfield  {journal} {\bibinfo  {journal} {arXiv:hep-th/9401072}\ }\href {https://doi.org/10.1016/0370-2693(94)91007-3} {10.1016/0370-2693(94)91007-3} (\bibinfo {year} {1994}),\ \Eprint {https://arxiv.org/abs/hep-th/9401072} {arXiv:hep-th/9401072} \BibitemShut {NoStop}%
\bibitem [{\citenamefont {Vidal}\ \emph {et~al.}(2003)\citenamefont {Vidal}, \citenamefont {Latorre}, \citenamefont {Rico},\ and\ \citenamefont {Kitaev}}]{vidal2003a}%
  \BibitemOpen
  \bibfield  {author} {\bibinfo {author} {\bibfnamefont {G.}~\bibnamefont {Vidal}}, \bibinfo {author} {\bibfnamefont {J.~I.}\ \bibnamefont {Latorre}}, \bibinfo {author} {\bibfnamefont {E.}~\bibnamefont {Rico}},\ and\ \bibinfo {author} {\bibfnamefont {A.}~\bibnamefont {Kitaev}},\ }\bibfield  {title} {\bibinfo {title} {Entanglement in {{Quantum Critical Phenomena}}},\ }\href {https://doi.org/10.1103/PhysRevLett.90.227902} {\bibfield  {journal} {\bibinfo  {journal} {Physical Review Letters}\ }\textbf {\bibinfo {volume} {90}},\ \bibinfo {pages} {227902} (\bibinfo {year} {2003})}\BibitemShut {NoStop}%
\bibitem [{\citenamefont {Calabrese}\ and\ \citenamefont {Cardy}(2004)}]{calabrese_2004}%
  \BibitemOpen
  \bibfield  {author} {\bibinfo {author} {\bibfnamefont {P.}~\bibnamefont {Calabrese}}\ and\ \bibinfo {author} {\bibfnamefont {J.}~\bibnamefont {Cardy}},\ }\bibfield  {title} {\bibinfo {title} {Entanglement entropy and quantum field theory},\ }\href {https://doi.org/DOI: 10.1088/1742-5468/2004/06/P06002; eprintid: arXiv:hep-th/0405152} {\bibfield  {journal} {\bibinfo  {journal} {Journal of Statistical Mechanics: Theory and Experiment}\ }\textbf {\bibinfo {volume} {06}},\ \bibinfo {pages} {002} (\bibinfo {year} {2004})}\BibitemShut {NoStop}%
\bibitem [{\citenamefont {Kitaev}\ and\ \citenamefont {Preskill}(2006)}]{kitaev2006}%
  \BibitemOpen
  \bibfield  {author} {\bibinfo {author} {\bibfnamefont {A.}~\bibnamefont {Kitaev}}\ and\ \bibinfo {author} {\bibfnamefont {J.}~\bibnamefont {Preskill}},\ }\bibfield  {title} {\bibinfo {title} {Topological {{Entanglement Entropy}}},\ }\href {https://doi.org/10.1103/PhysRevLett.96.110404} {\bibfield  {journal} {\bibinfo  {journal} {Physical Review Letters}\ }\textbf {\bibinfo {volume} {96}},\ \bibinfo {pages} {110404} (\bibinfo {year} {2006})}\BibitemShut {NoStop}%
\bibitem [{\citenamefont {Levin}\ and\ \citenamefont {Wen}(2006)}]{levin2006}%
  \BibitemOpen
  \bibfield  {author} {\bibinfo {author} {\bibfnamefont {M.}~\bibnamefont {Levin}}\ and\ \bibinfo {author} {\bibfnamefont {X.-G.}\ \bibnamefont {Wen}},\ }\bibfield  {title} {\bibinfo {title} {Detecting {{Topological Order}} in a {{Ground State Wave Function}}},\ }\href {https://doi.org/10.1103/PhysRevLett.96.110405} {\bibfield  {journal} {\bibinfo  {journal} {Physical Review Letters}\ }\textbf {\bibinfo {volume} {96}},\ \bibinfo {pages} {110405} (\bibinfo {year} {2006})}\BibitemShut {NoStop}%
\bibitem [{\citenamefont {Calabrese}\ and\ \citenamefont {Cardy}(2005)}]{calabrese_2005}%
  \BibitemOpen
  \bibfield  {author} {\bibinfo {author} {\bibfnamefont {P.}~\bibnamefont {Calabrese}}\ and\ \bibinfo {author} {\bibfnamefont {J.}~\bibnamefont {Cardy}},\ }\bibfield  {title} {\bibinfo {title} {Evolution of entanglement entropy in one-dimensional systems},\ }\href {https://doi.org/10.1088/1742-5468/2005/04/P04010} {\bibfield  {journal} {\bibinfo  {journal} {Journal of Statistical Mechanics: Theory and Experiment}\ }\textbf {\bibinfo {volume} {2005}},\ \bibinfo {pages} {P04010} (\bibinfo {year} {2005})}\BibitemShut {NoStop}%
\bibitem [{\citenamefont {Bardarson}\ \emph {et~al.}(2012)\citenamefont {Bardarson}, \citenamefont {Pollmann},\ and\ \citenamefont {Moore}}]{bardarson2012}%
  \BibitemOpen
  \bibfield  {author} {\bibinfo {author} {\bibfnamefont {J.~H.}\ \bibnamefont {Bardarson}}, \bibinfo {author} {\bibfnamefont {F.}~\bibnamefont {Pollmann}},\ and\ \bibinfo {author} {\bibfnamefont {J.~E.}\ \bibnamefont {Moore}},\ }\bibfield  {title} {\bibinfo {title} {Unbounded {{Growth}} of {{Entanglement}} in {{Models}} of {{Many-Body Localization}}},\ }\href {https://doi.org/10.1103/PhysRevLett.109.017202} {\bibfield  {journal} {\bibinfo  {journal} {Physical Review Letters}\ }\textbf {\bibinfo {volume} {109}},\ \bibinfo {pages} {017202} (\bibinfo {year} {2012})}\BibitemShut {NoStop}%
\bibitem [{\citenamefont {Serbyn}\ \emph {et~al.}(2013)\citenamefont {Serbyn}, \citenamefont {Papi{\'c}},\ and\ \citenamefont {Abanin}}]{serbyn2013}%
  \BibitemOpen
  \bibfield  {author} {\bibinfo {author} {\bibfnamefont {M.}~\bibnamefont {Serbyn}}, \bibinfo {author} {\bibfnamefont {Z.}~\bibnamefont {Papi{\'c}}},\ and\ \bibinfo {author} {\bibfnamefont {D.~A.}\ \bibnamefont {Abanin}},\ }\bibfield  {title} {\bibinfo {title} {Universal {{Slow Growth}} of {{Entanglement}} in {{Interacting Strongly Disordered Systems}}},\ }\href {https://doi.org/10.1103/PhysRevLett.110.260601} {\bibfield  {journal} {\bibinfo  {journal} {Physical Review Letters}\ }\textbf {\bibinfo {volume} {110}},\ \bibinfo {pages} {260601} (\bibinfo {year} {2013})}\BibitemShut {NoStop}%
\bibitem [{\citenamefont {Cardy}(2011)}]{cardy2011}%
  \BibitemOpen
  \bibfield  {author} {\bibinfo {author} {\bibfnamefont {J.}~\bibnamefont {Cardy}},\ }\bibfield  {title} {\bibinfo {title} {Measuring {{Entanglement Using Quantum Quenches}}},\ }\href {https://doi.org/10.1103/PhysRevLett.106.150404} {\bibfield  {journal} {\bibinfo  {journal} {Physical Review Letters}\ }\textbf {\bibinfo {volume} {106}},\ \bibinfo {pages} {150404} (\bibinfo {year} {2011})}\BibitemShut {NoStop}%
\bibitem [{\citenamefont {Abanin}\ and\ \citenamefont {Demler}(2012)}]{abanin2012}%
  \BibitemOpen
  \bibfield  {author} {\bibinfo {author} {\bibfnamefont {D.~A.}\ \bibnamefont {Abanin}}\ and\ \bibinfo {author} {\bibfnamefont {E.}~\bibnamefont {Demler}},\ }\bibfield  {title} {\bibinfo {title} {Measuring {{Entanglement Entropy}} of a {{Generic Many-Body System}} with a {{Quantum Switch}}},\ }\href {https://doi.org/10.1103/PhysRevLett.109.020504} {\bibfield  {journal} {\bibinfo  {journal} {Physical Review Letters}\ }\textbf {\bibinfo {volume} {109}},\ \bibinfo {pages} {020504} (\bibinfo {year} {2012})}\BibitemShut {NoStop}%
\bibitem [{\citenamefont {Daley}\ \emph {et~al.}(2012)\citenamefont {Daley}, \citenamefont {Pichler}, \citenamefont {Schachenmayer},\ and\ \citenamefont {Zoller}}]{daley2012}%
  \BibitemOpen
  \bibfield  {author} {\bibinfo {author} {\bibfnamefont {A.~J.}\ \bibnamefont {Daley}}, \bibinfo {author} {\bibfnamefont {H.}~\bibnamefont {Pichler}}, \bibinfo {author} {\bibfnamefont {J.}~\bibnamefont {Schachenmayer}},\ and\ \bibinfo {author} {\bibfnamefont {P.}~\bibnamefont {Zoller}},\ }\bibfield  {title} {\bibinfo {title} {Measuring {{Entanglement Growth}} in {{Quench Dynamics}} of {{Bosons}} in an {{Optical Lattice}}},\ }\href {https://doi.org/10.1103/PhysRevLett.109.020505} {\bibfield  {journal} {\bibinfo  {journal} {Physical Review Letters}\ }\textbf {\bibinfo {volume} {109}},\ \bibinfo {pages} {020505} (\bibinfo {year} {2012})}\BibitemShut {NoStop}%
\bibitem [{\citenamefont {Islam}\ \emph {et~al.}(2015)\citenamefont {Islam}, \citenamefont {Ma}, \citenamefont {Preiss}, \citenamefont {Tai}, \citenamefont {Lukin}, \citenamefont {Rispoli},\ and\ \citenamefont {Greiner}}]{islam2015}%
  \BibitemOpen
  \bibfield  {author} {\bibinfo {author} {\bibfnamefont {R.}~\bibnamefont {Islam}}, \bibinfo {author} {\bibfnamefont {R.}~\bibnamefont {Ma}}, \bibinfo {author} {\bibfnamefont {P.~M.}\ \bibnamefont {Preiss}}, \bibinfo {author} {\bibfnamefont {M.~E.}\ \bibnamefont {Tai}}, \bibinfo {author} {\bibfnamefont {A.}~\bibnamefont {Lukin}}, \bibinfo {author} {\bibfnamefont {M.}~\bibnamefont {Rispoli}},\ and\ \bibinfo {author} {\bibfnamefont {M.}~\bibnamefont {Greiner}},\ }\bibfield  {title} {\bibinfo {title} {Measuring entanglement entropy in a quantum many-body system},\ }\href {https://doi.org/10.1038/nature15750} {\bibfield  {journal} {\bibinfo  {journal} {Nature}\ }\textbf {\bibinfo {volume} {528}},\ \bibinfo {pages} {77} (\bibinfo {year} {2015})}\BibitemShut {NoStop}%
\bibitem [{\citenamefont {Kaufman}\ \emph {et~al.}(2016)\citenamefont {Kaufman}, \citenamefont {Tai}, \citenamefont {Lukin}, \citenamefont {Rispoli}, \citenamefont {Schittko}, \citenamefont {Preiss},\ and\ \citenamefont {Greiner}}]{kaufman2016}%
  \BibitemOpen
  \bibfield  {author} {\bibinfo {author} {\bibfnamefont {A.~M.}\ \bibnamefont {Kaufman}}, \bibinfo {author} {\bibfnamefont {M.~E.}\ \bibnamefont {Tai}}, \bibinfo {author} {\bibfnamefont {A.}~\bibnamefont {Lukin}}, \bibinfo {author} {\bibfnamefont {M.}~\bibnamefont {Rispoli}}, \bibinfo {author} {\bibfnamefont {R.}~\bibnamefont {Schittko}}, \bibinfo {author} {\bibfnamefont {P.~M.}\ \bibnamefont {Preiss}},\ and\ \bibinfo {author} {\bibfnamefont {M.}~\bibnamefont {Greiner}},\ }\bibfield  {title} {\bibinfo {title} {Quantum thermalization through entanglement in an isolated many-body system},\ }\href {https://doi.org/10.1126/science.aaf6725} {\bibfield  {journal} {\bibinfo  {journal} {Science}\ }\textbf {\bibinfo {volume} {353}},\ \bibinfo {pages} {794} (\bibinfo {year} {2016})}\BibitemShut {NoStop}%
\bibitem [{\citenamefont {Hauke}\ \emph {et~al.}(2016)\citenamefont {Hauke}, \citenamefont {Heyl}, \citenamefont {Tagliacozzo},\ and\ \citenamefont {Zoller}}]{hauke_2016}%
  \BibitemOpen
  \bibfield  {author} {\bibinfo {author} {\bibfnamefont {P.}~\bibnamefont {Hauke}}, \bibinfo {author} {\bibfnamefont {M.}~\bibnamefont {Heyl}}, \bibinfo {author} {\bibfnamefont {L.}~\bibnamefont {Tagliacozzo}},\ and\ \bibinfo {author} {\bibfnamefont {P.}~\bibnamefont {Zoller}},\ }\bibfield  {title} {\bibinfo {title} {Measuring multipartite entanglement through dynamic susceptibilities},\ }\href {https://doi.org/10.1038/nphys3700} {\bibfield  {journal} {\bibinfo  {journal} {Nature Physics}\ }\textbf {\bibinfo {volume} {12}},\ \bibinfo {pages} {778} (\bibinfo {year} {2016})}\BibitemShut {NoStop}%
\bibitem [{\citenamefont {Brydges}\ \emph {et~al.}(2019)\citenamefont {Brydges}, \citenamefont {Elben}, \citenamefont {Jurcevic}, \citenamefont {Vermersch}, \citenamefont {Maier}, \citenamefont {Lanyon}, \citenamefont {Zoller}, \citenamefont {Blatt},\ and\ \citenamefont {Roos}}]{brydges2019}%
  \BibitemOpen
  \bibfield  {author} {\bibinfo {author} {\bibfnamefont {T.}~\bibnamefont {Brydges}}, \bibinfo {author} {\bibfnamefont {A.}~\bibnamefont {Elben}}, \bibinfo {author} {\bibfnamefont {P.}~\bibnamefont {Jurcevic}}, \bibinfo {author} {\bibfnamefont {B.}~\bibnamefont {Vermersch}}, \bibinfo {author} {\bibfnamefont {C.}~\bibnamefont {Maier}}, \bibinfo {author} {\bibfnamefont {B.~P.}\ \bibnamefont {Lanyon}}, \bibinfo {author} {\bibfnamefont {P.}~\bibnamefont {Zoller}}, \bibinfo {author} {\bibfnamefont {R.}~\bibnamefont {Blatt}},\ and\ \bibinfo {author} {\bibfnamefont {C.~F.}\ \bibnamefont {Roos}},\ }\bibfield  {title} {\bibinfo {title} {Probing {{R{\'e}nyi}} entanglement entropy via randomized measurements},\ }\href {https://doi.org/10.1126/science.aau4963} {\bibfield  {journal} {\bibinfo  {journal} {Science}\ }\textbf {\bibinfo {volume} {364}},\ \bibinfo {pages} {260} (\bibinfo {year} {2019})}\BibitemShut {NoStop}%
\bibitem [{\citenamefont {Huang}\ \emph {et~al.}(2020)\citenamefont {Huang}, \citenamefont {Kueng},\ and\ \citenamefont {Preskill}}]{huang2020}%
  \BibitemOpen
  \bibfield  {author} {\bibinfo {author} {\bibfnamefont {H.-Y.}\ \bibnamefont {Huang}}, \bibinfo {author} {\bibfnamefont {R.}~\bibnamefont {Kueng}},\ and\ \bibinfo {author} {\bibfnamefont {J.}~\bibnamefont {Preskill}},\ }\bibfield  {title} {\bibinfo {title} {Predicting many properties of a quantum system from very few measurements},\ }\href {https://doi.org/10.1038/s41567-020-0932-7} {\bibfield  {journal} {\bibinfo  {journal} {Nature Physics}\ }\textbf {\bibinfo {volume} {16}},\ \bibinfo {pages} {1050} (\bibinfo {year} {2020})}\BibitemShut {NoStop}%
\bibitem [{\citenamefont {Elben}\ \emph {et~al.}(2020)\citenamefont {Elben}, \citenamefont {Kueng}, \citenamefont {Huang}, \citenamefont {{van Bijnen}}, \citenamefont {Kokail}, \citenamefont {Dalmonte}, \citenamefont {Calabrese}, \citenamefont {Kraus}, \citenamefont {Preskill}, \citenamefont {Zoller},\ and\ \citenamefont {Vermersch}}]{elben2020}%
  \BibitemOpen
  \bibfield  {author} {\bibinfo {author} {\bibfnamefont {A.}~\bibnamefont {Elben}}, \bibinfo {author} {\bibfnamefont {R.}~\bibnamefont {Kueng}}, \bibinfo {author} {\bibfnamefont {H.-Y.}\ \bibnamefont {Huang}}, \bibinfo {author} {\bibfnamefont {R.}~\bibnamefont {{van Bijnen}}}, \bibinfo {author} {\bibfnamefont {C.}~\bibnamefont {Kokail}}, \bibinfo {author} {\bibfnamefont {M.}~\bibnamefont {Dalmonte}}, \bibinfo {author} {\bibfnamefont {P.}~\bibnamefont {Calabrese}}, \bibinfo {author} {\bibfnamefont {B.}~\bibnamefont {Kraus}}, \bibinfo {author} {\bibfnamefont {J.}~\bibnamefont {Preskill}}, \bibinfo {author} {\bibfnamefont {P.}~\bibnamefont {Zoller}},\ and\ \bibinfo {author} {\bibfnamefont {B.}~\bibnamefont {Vermersch}},\ }\href {https://doi.org/10.1103/PhysRevLett.125.200501} {\bibinfo {title} {Mixed-state entanglement from local randomized measurements}} (\bibinfo {year} {2020}),\ \Eprint {https://arxiv.org/abs/2007.06305} {arXiv:2007.06305 [cond-mat, physics:quant-ph]} \BibitemShut {NoStop}%
\bibitem [{\citenamefont {Leifer}(2006)}]{leifer2006}%
  \BibitemOpen
  \bibfield  {author} {\bibinfo {author} {\bibfnamefont {M.~S.}\ \bibnamefont {Leifer}},\ }\bibfield  {title} {\bibinfo {title} {Quantum dynamics as an analog of conditional probability},\ }\bibfield  {journal} {\bibinfo  {journal} {Physical Review A}\ }\textbf {\bibinfo {volume} {74}},\ \href {https://doi.org/10.1103/PhysRevA.74.042310} {10.1103/PhysRevA.74.042310} (\bibinfo {year} {2006})\BibitemShut {NoStop}%
\bibitem [{\citenamefont {Leifer}\ and\ \citenamefont {Spekkens}(2013)}]{leifer2013}%
  \BibitemOpen
  \bibfield  {author} {\bibinfo {author} {\bibfnamefont {M.~S.}\ \bibnamefont {Leifer}}\ and\ \bibinfo {author} {\bibfnamefont {R.~W.}\ \bibnamefont {Spekkens}},\ }\bibfield  {title} {\bibinfo {title} {Towards a formulation of quantum theory as a causally neutral theory of {{Bayesian}} inference},\ }\href {https://doi.org/10.1103/PhysRevA.88.052130} {\bibfield  {journal} {\bibinfo  {journal} {Physical Review A}\ }\textbf {\bibinfo {volume} {88}},\ \bibinfo {pages} {052130} (\bibinfo {year} {2013})}\BibitemShut {NoStop}%
\bibitem [{\citenamefont {Parzygnat}\ and\ \citenamefont {Fullwood}(2023)}]{parzygnat2023}%
  \BibitemOpen
  \bibfield  {author} {\bibinfo {author} {\bibfnamefont {A.~J.}\ \bibnamefont {Parzygnat}}\ and\ \bibinfo {author} {\bibfnamefont {J.}~\bibnamefont {Fullwood}},\ }\bibfield  {title} {\bibinfo {title} {From {{Time-Reversal Symmetry}} to {{Quantum Bayes}}' {{Rules}}},\ }\href {https://doi.org/10.1103/PRXQuantum.4.020334} {\bibfield  {journal} {\bibinfo  {journal} {PRX Quantum}\ }\textbf {\bibinfo {volume} {4}},\ \bibinfo {pages} {020334} (\bibinfo {year} {2023})}\BibitemShut {NoStop}%
\bibitem [{\citenamefont {Feynman}\ and\ \citenamefont {Vernon}(1963)}]{feynman1963}%
  \BibitemOpen
  \bibfield  {author} {\bibinfo {author} {\bibfnamefont {R.}~\bibnamefont {Feynman}}\ and\ \bibinfo {author} {\bibfnamefont {F.}~\bibnamefont {Vernon}},\ }\bibfield  {title} {\bibinfo {title} {The theory of a general quantum system interacting with a linear dissipative system},\ }\href {https://doi.org/10.1016/0003-4916(63)90068-X} {\bibfield  {journal} {\bibinfo  {journal} {Annals of Physics}\ }\textbf {\bibinfo {volume} {24}},\ \bibinfo {pages} {118} (\bibinfo {year} {1963})}\BibitemShut {NoStop}%
\bibitem [{\citenamefont {Ba{\~n}uls}\ \emph {et~al.}(2009)\citenamefont {Ba{\~n}uls}, \citenamefont {Hastings}, \citenamefont {Verstraete},\ and\ \citenamefont {Cirac}}]{banuls2009}%
  \BibitemOpen
  \bibfield  {author} {\bibinfo {author} {\bibfnamefont {M.~C.}\ \bibnamefont {Ba{\~n}uls}}, \bibinfo {author} {\bibfnamefont {M.~B.}\ \bibnamefont {Hastings}}, \bibinfo {author} {\bibfnamefont {F.}~\bibnamefont {Verstraete}},\ and\ \bibinfo {author} {\bibfnamefont {J.~I.}\ \bibnamefont {Cirac}},\ }\bibfield  {title} {\bibinfo {title} {Matrix {{Product States}} for {{Dynamical Simulation}} of {{Infinite Chains}}},\ }\href {https://doi.org/10.1103/PhysRevLett.102.240603} {\bibfield  {journal} {\bibinfo  {journal} {Physical Review Letters}\ }\textbf {\bibinfo {volume} {102}},\ \bibinfo {pages} {240603} (\bibinfo {year} {2009})}\BibitemShut {NoStop}%
\bibitem [{\citenamefont {{M{\"u}ller-Hermes}}\ \emph {et~al.}(2012)\citenamefont {{M{\"u}ller-Hermes}}, \citenamefont {Cirac},\ and\ \citenamefont {Ba{\~n}uls}}]{muller-hermes2012}%
  \BibitemOpen
  \bibfield  {author} {\bibinfo {author} {\bibfnamefont {A.}~\bibnamefont {{M{\"u}ller-Hermes}}}, \bibinfo {author} {\bibfnamefont {J.~I.}\ \bibnamefont {Cirac}},\ and\ \bibinfo {author} {\bibfnamefont {M.~C.}\ \bibnamefont {Ba{\~n}uls}},\ }\bibfield  {title} {\bibinfo {title} {Tensor network techniques for the computation of dynamical observables in {{1D}} quantum spin systems},\ }\href {https://doi.org/10.1088/1367-2630/14/7/075003} {\bibfield  {journal} {\bibinfo  {journal} {New Journal of Physics}\ }\textbf {\bibinfo {volume} {14}},\ \bibinfo {pages} {075003} (\bibinfo {year} {2012})},\ \Eprint {https://arxiv.org/abs/1204.5080} {arXiv:1204.5080 [cond-mat, physics:quant-ph]} \BibitemShut {NoStop}%
\bibitem [{\citenamefont {Hastings}\ and\ \citenamefont {Mahajan}(2015)}]{hastings2015}%
  \BibitemOpen
  \bibfield  {author} {\bibinfo {author} {\bibfnamefont {M.~B.}\ \bibnamefont {Hastings}}\ and\ \bibinfo {author} {\bibfnamefont {R.}~\bibnamefont {Mahajan}},\ }\bibfield  {title} {\bibinfo {title} {Connecting {{Entanglement}} in {{Time}} and {{Space}}: {{Improving}} the {{Folding Algorithm}}},\ }\href {https://doi.org/10.1103/PhysRevA.91.032306} {\bibfield  {journal} {\bibinfo  {journal} {Physical Review A}\ }\textbf {\bibinfo {volume} {91}},\ \bibinfo {pages} {032306} (\bibinfo {year} {2015})},\ \Eprint {https://arxiv.org/abs/1411.7950} {arXiv:1411.7950 [cond-mat, physics:hep-th, physics:quant-ph]} \BibitemShut {NoStop}%
\bibitem [{\citenamefont {Lerose}\ \emph {et~al.}(2021)\citenamefont {Lerose}, \citenamefont {Sonner},\ and\ \citenamefont {Abanin}}]{lerose2021}%
  \BibitemOpen
  \bibfield  {author} {\bibinfo {author} {\bibfnamefont {A.}~\bibnamefont {Lerose}}, \bibinfo {author} {\bibfnamefont {M.}~\bibnamefont {Sonner}},\ and\ \bibinfo {author} {\bibfnamefont {D.~A.}\ \bibnamefont {Abanin}},\ }\bibfield  {title} {\bibinfo {title} {Influence {{Matrix Approach}} to {{Many-Body Floquet Dynamics}}},\ }\href {https://doi.org/10.1103/PhysRevX.11.021040} {\bibfield  {journal} {\bibinfo  {journal} {Physical Review X}\ }\textbf {\bibinfo {volume} {11}},\ \bibinfo {pages} {021040} (\bibinfo {year} {2021})}\BibitemShut {NoStop}%
\bibitem [{\citenamefont {Sonner}\ \emph {et~al.}(2021)\citenamefont {Sonner}, \citenamefont {Lerose},\ and\ \citenamefont {Abanin}}]{sonner2021}%
  \BibitemOpen
  \bibfield  {author} {\bibinfo {author} {\bibfnamefont {M.}~\bibnamefont {Sonner}}, \bibinfo {author} {\bibfnamefont {A.}~\bibnamefont {Lerose}},\ and\ \bibinfo {author} {\bibfnamefont {D.~A.}\ \bibnamefont {Abanin}},\ }\bibfield  {title} {\bibinfo {title} {Influence functional of many-body systems: {{Temporal}} entanglement and matrix-product state representation},\ }\href {https://doi.org/10.1016/j.aop.2021.168677} {\bibfield  {journal} {\bibinfo  {journal} {Annals of Physics}\ }\bibinfo {series} {Special Issue on {{Philip W}}. {{Anderson}}},\ \textbf {\bibinfo {volume} {435}},\ \bibinfo {pages} {168677} (\bibinfo {year} {2021})}\BibitemShut {NoStop}%
\bibitem [{\citenamefont {Giudice}\ \emph {et~al.}(2022)\citenamefont {Giudice}, \citenamefont {Giudici}, \citenamefont {Sonner}, \citenamefont {Thoenniss}, \citenamefont {Lerose}, \citenamefont {Abanin},\ and\ \citenamefont {Piroli}}]{giudice2022}%
  \BibitemOpen
  \bibfield  {author} {\bibinfo {author} {\bibfnamefont {G.}~\bibnamefont {Giudice}}, \bibinfo {author} {\bibfnamefont {G.}~\bibnamefont {Giudici}}, \bibinfo {author} {\bibfnamefont {M.}~\bibnamefont {Sonner}}, \bibinfo {author} {\bibfnamefont {J.}~\bibnamefont {Thoenniss}}, \bibinfo {author} {\bibfnamefont {A.}~\bibnamefont {Lerose}}, \bibinfo {author} {\bibfnamefont {D.~A.}\ \bibnamefont {Abanin}},\ and\ \bibinfo {author} {\bibfnamefont {L.}~\bibnamefont {Piroli}},\ }\bibfield  {title} {\bibinfo {title} {Temporal {{Entanglement}}, {{Quasiparticles}}, and the {{Role}} of {{Interactions}}},\ }\href {https://doi.org/10.1103/PhysRevLett.128.220401} {\bibfield  {journal} {\bibinfo  {journal} {Physical Review Letters}\ }\textbf {\bibinfo {volume} {128}},\ \bibinfo {pages} {220401} (\bibinfo {year} {2022})}\BibitemShut {NoStop}%
\bibitem [{\citenamefont {Carignano}\ \emph {et~al.}(2024)\citenamefont {Carignano}, \citenamefont {Marim{\'o}n},\ and\ \citenamefont {Tagliacozzo}}]{carignano2024a}%
  \BibitemOpen
  \bibfield  {author} {\bibinfo {author} {\bibfnamefont {S.}~\bibnamefont {Carignano}}, \bibinfo {author} {\bibfnamefont {C.~R.}\ \bibnamefont {Marim{\'o}n}},\ and\ \bibinfo {author} {\bibfnamefont {L.}~\bibnamefont {Tagliacozzo}},\ }\bibfield  {title} {\bibinfo {title} {Temporal entropy and the complexity of computing the expectation value of local operators after a quench},\ }\href {https://doi.org/10.1103/PhysRevResearch.6.033021} {\bibfield  {journal} {\bibinfo  {journal} {Physical Review Research}\ }\textbf {\bibinfo {volume} {6}},\ \bibinfo {pages} {033021} (\bibinfo {year} {2024})}\BibitemShut {NoStop}%
\bibitem [{\citenamefont {Guo}\ \emph {et~al.}(2024{\natexlab{a}})\citenamefont {Guo}, \citenamefont {He},\ and\ \citenamefont {Zhang}}]{guo2024a}%
  \BibitemOpen
  \bibfield  {author} {\bibinfo {author} {\bibfnamefont {W.-z.}\ \bibnamefont {Guo}}, \bibinfo {author} {\bibfnamefont {S.}~\bibnamefont {He}},\ and\ \bibinfo {author} {\bibfnamefont {Y.-X.}\ \bibnamefont {Zhang}},\ }\href {https://doi.org/10.48550/arXiv.2402.00268} {\bibinfo {title} {Relation between timelike and spacelike entanglement entropy}} (\bibinfo {year} {2024}{\natexlab{a}}),\ \Eprint {https://arxiv.org/abs/2402.00268} {arXiv:2402.00268 [cond-mat, physics:hep-th, physics:quant-ph]} \BibitemShut {NoStop}%
\bibitem [{\citenamefont {Yao}\ and\ \citenamefont {Claeys}(2024)}]{yao2024a}%
  \BibitemOpen
  \bibfield  {author} {\bibinfo {author} {\bibfnamefont {J.}~\bibnamefont {Yao}}\ and\ \bibinfo {author} {\bibfnamefont {P.~W.}\ \bibnamefont {Claeys}},\ }\href {https://doi.org/10.48550/arXiv.2404.14374} {\bibinfo {title} {Temporal {{Entanglement Profiles}} in {{Dual-Unitary Clifford Circuits}} with {{Measurements}}}} (\bibinfo {year} {2024}),\ \Eprint {https://arxiv.org/abs/2404.14374} {arXiv:2404.14374 [cond-mat, physics:quant-ph]} \BibitemShut {NoStop}%
\bibitem [{\citenamefont {Foligno}\ \emph {et~al.}(2023)\citenamefont {Foligno}, \citenamefont {Zhou},\ and\ \citenamefont {Bertini}}]{foligno2023a}%
  \BibitemOpen
  \bibfield  {author} {\bibinfo {author} {\bibfnamefont {A.}~\bibnamefont {Foligno}}, \bibinfo {author} {\bibfnamefont {T.}~\bibnamefont {Zhou}},\ and\ \bibinfo {author} {\bibfnamefont {B.}~\bibnamefont {Bertini}},\ }\bibfield  {title} {\bibinfo {title} {Temporal {{Entanglement}} in {{Chaotic Quantum Circuits}}},\ }\href {https://doi.org/10.1103/PhysRevX.13.041008} {\bibfield  {journal} {\bibinfo  {journal} {Physical Review X}\ }\textbf {\bibinfo {volume} {13}},\ \bibinfo {pages} {041008} (\bibinfo {year} {2023})},\ \Eprint {https://arxiv.org/abs/2302.08502} {arXiv:2302.08502 [cond-mat, physics:hep-th, physics:math-ph, physics:quant-ph]} \BibitemShut {NoStop}%
\bibitem [{\citenamefont {Nakata}\ \emph {et~al.}(2021)\citenamefont {Nakata}, \citenamefont {Takayanagi}, \citenamefont {Taki}, \citenamefont {Tamaoka},\ and\ \citenamefont {Wei}}]{nakata2021}%
  \BibitemOpen
  \bibfield  {author} {\bibinfo {author} {\bibfnamefont {Y.}~\bibnamefont {Nakata}}, \bibinfo {author} {\bibfnamefont {T.}~\bibnamefont {Takayanagi}}, \bibinfo {author} {\bibfnamefont {Y.}~\bibnamefont {Taki}}, \bibinfo {author} {\bibfnamefont {K.}~\bibnamefont {Tamaoka}},\ and\ \bibinfo {author} {\bibfnamefont {Z.}~\bibnamefont {Wei}},\ }\bibfield  {title} {\bibinfo {title} {Holographic {{Pseudo Entropy}}},\ }\href {https://doi.org/10.1103/PhysRevD.103.026005} {\bibfield  {journal} {\bibinfo  {journal} {Physical Review D}\ }\textbf {\bibinfo {volume} {103}},\ \bibinfo {pages} {026005} (\bibinfo {year} {2021})},\ \Eprint {https://arxiv.org/abs/2005.13801} {arXiv:2005.13801 [cond-mat, physics:hep-th, physics:quant-ph]} \BibitemShut {NoStop}%
\bibitem [{\citenamefont {Doi}\ \emph {et~al.}(2023{\natexlab{a}})\citenamefont {Doi}, \citenamefont {Harper}, \citenamefont {Mollabashi}, \citenamefont {Takayanagi},\ and\ \citenamefont {Taki}}]{doi2023a}%
  \BibitemOpen
  \bibfield  {author} {\bibinfo {author} {\bibfnamefont {K.}~\bibnamefont {Doi}}, \bibinfo {author} {\bibfnamefont {J.}~\bibnamefont {Harper}}, \bibinfo {author} {\bibfnamefont {A.}~\bibnamefont {Mollabashi}}, \bibinfo {author} {\bibfnamefont {T.}~\bibnamefont {Takayanagi}},\ and\ \bibinfo {author} {\bibfnamefont {Y.}~\bibnamefont {Taki}},\ }\bibfield  {title} {\bibinfo {title} {Pseudo {{Entropy}} in {{dS}}/{{CFT}} and {{Time-like Entanglement Entropy}}},\ }\href {https://doi.org/10.1103/PhysRevLett.130.031601} {\bibfield  {journal} {\bibinfo  {journal} {Physical Review Letters}\ }\textbf {\bibinfo {volume} {130}},\ \bibinfo {pages} {031601} (\bibinfo {year} {2023}{\natexlab{a}})},\ \Eprint {https://arxiv.org/abs/2210.09457} {arXiv:2210.09457 [cond-mat, physics:hep-th, physics:quant-ph]} \BibitemShut {NoStop}%
\bibitem [{\citenamefont {Doi}\ \emph {et~al.}(2023{\natexlab{b}})\citenamefont {Doi}, \citenamefont {Harper}, \citenamefont {Mollabashi}, \citenamefont {Takayanagi},\ and\ \citenamefont {Taki}}]{doi2023b}%
  \BibitemOpen
  \bibfield  {author} {\bibinfo {author} {\bibfnamefont {K.}~\bibnamefont {Doi}}, \bibinfo {author} {\bibfnamefont {J.}~\bibnamefont {Harper}}, \bibinfo {author} {\bibfnamefont {A.}~\bibnamefont {Mollabashi}}, \bibinfo {author} {\bibfnamefont {T.}~\bibnamefont {Takayanagi}},\ and\ \bibinfo {author} {\bibfnamefont {Y.}~\bibnamefont {Taki}},\ }\bibfield  {title} {\bibinfo {title} {Timelike entanglement entropy},\ }\href {https://doi.org/10.1007/JHEP05(2023)052} {\bibfield  {journal} {\bibinfo  {journal} {Journal of High Energy Physics}\ }\textbf {\bibinfo {volume} {2023}},\ \bibinfo {pages} {52} (\bibinfo {year} {2023}{\natexlab{b}})}\BibitemShut {NoStop}%
\bibitem [{\citenamefont {Narayan}\ and\ \citenamefont {Saini}(2023)}]{narayan2023}%
  \BibitemOpen
  \bibfield  {author} {\bibinfo {author} {\bibfnamefont {K.}~\bibnamefont {Narayan}}\ and\ \bibinfo {author} {\bibfnamefont {H.~K.}\ \bibnamefont {Saini}},\ }\href {https://doi.org/10.48550/arXiv.2303.01307} {\bibinfo {title} {Notes on time entanglement and pseudo-entropy}} (\bibinfo {year} {2023}),\ \Eprint {https://arxiv.org/abs/2303.01307} {arXiv:2303.01307 [hep-th]} \BibitemShut {NoStop}%
\bibitem [{\citenamefont {Narayan}(2023)}]{narayan2023b}%
  \BibitemOpen
  \bibfield  {author} {\bibinfo {author} {\bibfnamefont {K.}~\bibnamefont {Narayan}},\ }\bibfield  {title} {\bibinfo {title} {De {{Sitter}} space, extremal surfaces, and time entanglement},\ }\href {https://doi.org/10.1103/PhysRevD.107.126004} {\bibfield  {journal} {\bibinfo  {journal} {Physical Review D}\ }\textbf {\bibinfo {volume} {107}},\ \bibinfo {pages} {126004} (\bibinfo {year} {2023})}\BibitemShut {NoStop}%
\bibitem [{\citenamefont {Heller}\ \emph {et~al.}(2024)\citenamefont {Heller}, \citenamefont {Ori},\ and\ \citenamefont {Serantes}}]{heller2024}%
  \BibitemOpen
  \bibfield  {author} {\bibinfo {author} {\bibfnamefont {M.~P.}\ \bibnamefont {Heller}}, \bibinfo {author} {\bibfnamefont {F.}~\bibnamefont {Ori}},\ and\ \bibinfo {author} {\bibfnamefont {A.}~\bibnamefont {Serantes}},\ }\href {https://doi.org/10.48550/arXiv.2408.15752} {\bibinfo {title} {Geometric interpretation of timelike entanglement entropy}} (\bibinfo {year} {2024}),\ \Eprint {https://arxiv.org/abs/2408.15752} {arXiv:2408.15752 [gr-qc, physics:hep-th, physics:quant-ph]} \BibitemShut {NoStop}%
\bibitem [{\citenamefont {Mollabashi}\ \emph {et~al.}(2021)\citenamefont {Mollabashi}, \citenamefont {Shiba}, \citenamefont {Takayanagi}, \citenamefont {Tamaoka},\ and\ \citenamefont {Wei}}]{mollabashi2021}%
  \BibitemOpen
  \bibfield  {author} {\bibinfo {author} {\bibfnamefont {A.}~\bibnamefont {Mollabashi}}, \bibinfo {author} {\bibfnamefont {N.}~\bibnamefont {Shiba}}, \bibinfo {author} {\bibfnamefont {T.}~\bibnamefont {Takayanagi}}, \bibinfo {author} {\bibfnamefont {K.}~\bibnamefont {Tamaoka}},\ and\ \bibinfo {author} {\bibfnamefont {Z.}~\bibnamefont {Wei}},\ }\bibfield  {title} {\bibinfo {title} {Pseudo-{{Entropy}} in {{Free Quantum Field Theories}}},\ }\href {https://doi.org/10.1103/PhysRevLett.126.081601} {\bibfield  {journal} {\bibinfo  {journal} {Physical Review Letters}\ }\textbf {\bibinfo {volume} {126}},\ \bibinfo {pages} {081601} (\bibinfo {year} {2021})}\BibitemShut {NoStop}%
\bibitem [{\citenamefont {Murciano}\ \emph {et~al.}(2022)\citenamefont {Murciano}, \citenamefont {Calabrese},\ and\ \citenamefont {Konik}}]{murciano2022}%
  \BibitemOpen
  \bibfield  {author} {\bibinfo {author} {\bibfnamefont {S.}~\bibnamefont {Murciano}}, \bibinfo {author} {\bibfnamefont {P.}~\bibnamefont {Calabrese}},\ and\ \bibinfo {author} {\bibfnamefont {R.~M.}\ \bibnamefont {Konik}},\ }\bibfield  {title} {\bibinfo {title} {Generalized entanglement entropies in two-dimensional conformal field theory},\ }\href {https://doi.org/10.1007/JHEP05(2022)152} {\bibfield  {journal} {\bibinfo  {journal} {Journal of High Energy Physics}\ }\textbf {\bibinfo {volume} {2022}},\ \bibinfo {pages} {152} (\bibinfo {year} {2022})}\BibitemShut {NoStop}%
\bibitem [{\citenamefont {Vidal}(2003)}]{vidal2003}%
  \BibitemOpen
  \bibfield  {author} {\bibinfo {author} {\bibfnamefont {G.}~\bibnamefont {Vidal}},\ }\bibfield  {title} {\bibinfo {title} {Efficient classical simulation of slightly entangled quantum computations},\ }\href {https://doi.org/10.1103/PhysRevLett.91.147902} {\bibfield  {journal} {\bibinfo  {journal} {Physical Review Letters}\ }\textbf {\bibinfo {volume} {91}},\ \bibinfo {pages} {147902} (\bibinfo {year} {2003})}\BibitemShut {NoStop}%
\bibitem [{\citenamefont {Schollw{\"o}ck}(2011)}]{schollwock2011}%
  \BibitemOpen
  \bibfield  {author} {\bibinfo {author} {\bibfnamefont {U.}~\bibnamefont {Schollw{\"o}ck}},\ }\bibfield  {title} {\bibinfo {title} {The density-matrix renormalization group in the age of matrix product states},\ }\href {https://doi.org/10.1016/j.aop.2010.09.012} {\bibfield  {journal} {\bibinfo  {journal} {Ann. Phys. (N. Y).}\ }\textbf {\bibinfo {volume} {326}},\ \bibinfo {pages} {96} (\bibinfo {year} {2011})}\BibitemShut {NoStop}%
\bibitem [{\citenamefont {Paeckel}\ \emph {et~al.}(2019)\citenamefont {Paeckel}, \citenamefont {K{\"o}hler}, \citenamefont {Swoboda}, \citenamefont {Manmana}, \citenamefont {Schollw{\"o}ck},\ and\ \citenamefont {Hubig}}]{paeckel2019}%
  \BibitemOpen
  \bibfield  {author} {\bibinfo {author} {\bibfnamefont {S.}~\bibnamefont {Paeckel}}, \bibinfo {author} {\bibfnamefont {T.}~\bibnamefont {K{\"o}hler}}, \bibinfo {author} {\bibfnamefont {A.}~\bibnamefont {Swoboda}}, \bibinfo {author} {\bibfnamefont {S.~R.}\ \bibnamefont {Manmana}}, \bibinfo {author} {\bibfnamefont {U.}~\bibnamefont {Schollw{\"o}ck}},\ and\ \bibinfo {author} {\bibfnamefont {C.}~\bibnamefont {Hubig}},\ }\bibfield  {title} {\bibinfo {title} {Time-evolution methods for matrix-product states},\ }\href@noop {} {\bibfield  {journal} {\bibinfo  {journal} {arXiv:1901.05824 [cond-mat, physics:quant-ph]}\ } (\bibinfo {year} {2019})},\ \Eprint {https://arxiv.org/abs/1901.05824} {arXiv:1901.05824} \BibitemShut {NoStop}%
\bibitem [{\citenamefont {Carignano}\ and\ \citenamefont {Tagliacozzo}(2024)}]{carignano2024}%
  \BibitemOpen
  \bibfield  {author} {\bibinfo {author} {\bibfnamefont {S.}~\bibnamefont {Carignano}}\ and\ \bibinfo {author} {\bibfnamefont {L.}~\bibnamefont {Tagliacozzo}},\ }\href {https://doi.org/10.48550/arXiv.2405.14706} {\bibinfo {title} {Loschmidt echo, emerging dual unitarity and scaling of generalized temporal entropies after quenches to the critical point}} (\bibinfo {year} {2024}),\ \Eprint {https://arxiv.org/abs/2405.14706} {arXiv:2405.14706 [cond-mat]} \BibitemShut {NoStop}%
\bibitem [{\citenamefont {Cerezo-Roquebrún}\ \emph {et~al.}(2025)\citenamefont {Cerezo-Roquebrún}, \citenamefont {Bou-Comas}, \citenamefont {Schneider}, \citenamefont {López}, \citenamefont {Tagliacozzo},\ and\ \citenamefont {Carignano}}]{cerezo-roquebrún2025}%
  \BibitemOpen
  \bibfield  {author} {\bibinfo {author} {\bibfnamefont {S.}~\bibnamefont {Cerezo-Roquebrún}}, \bibinfo {author} {\bibfnamefont {A.}~\bibnamefont {Bou-Comas}}, \bibinfo {author} {\bibfnamefont {J.~T.}\ \bibnamefont {Schneider}}, \bibinfo {author} {\bibfnamefont {E.}~\bibnamefont {López}}, \bibinfo {author} {\bibfnamefont {L.}~\bibnamefont {Tagliacozzo}},\ and\ \bibinfo {author} {\bibfnamefont {S.}~\bibnamefont {Carignano}},\ }\bibfield  {title} {\bibinfo {title} {Spatio-temporal tensor-network approaches to out-of-equilibrium dynamics bridging open and closed systems},\ }\bibfield  {journal} {\bibinfo  {journal} {Frontiers in Quantum Science and Technology}\ }\textbf {\bibinfo {volume} {4}},\ \href {https://doi.org/10.3389/frqst.2025.1568471} {10.3389/frqst.2025.1568471} (\bibinfo {year} {2025})\BibitemShut {NoStop}%
\bibitem [{\citenamefont {Zanardi}(2001)}]{zanardi2001}%
  \BibitemOpen
  \bibfield  {author} {\bibinfo {author} {\bibfnamefont {P.}~\bibnamefont {Zanardi}},\ }\bibfield  {title} {\bibinfo {title} {Entanglement of quantum evolutions},\ }\href {https://doi.org/10.1103/PhysRevA.63.040304} {\bibfield  {journal} {\bibinfo  {journal} {Physical Review A}\ }\textbf {\bibinfo {volume} {63}},\ \bibinfo {pages} {040304} (\bibinfo {year} {2001})}\BibitemShut {NoStop}%
\bibitem [{\citenamefont {Dubail}(2017)}]{dubail2017}%
  \BibitemOpen
  \bibfield  {author} {\bibinfo {author} {\bibfnamefont {J.}~\bibnamefont {Dubail}},\ }\bibfield  {title} {\bibinfo {title} {Entanglement scaling of operators: A conformal field theory approach, with a glimpse of simulability of long-time dynamics in 1+1d},\ }\href {https://doi.org/10.1088/1751-8121/aa6f38} {\bibfield  {journal} {\bibinfo  {journal} {Journal of Physics A: Mathematical and Theoretical}\ }\textbf {\bibinfo {volume} {50}},\ \bibinfo {pages} {234001} (\bibinfo {year} {2017})},\ \Eprint {https://arxiv.org/abs/1612.08630} {arXiv:1612.08630 [cond-mat, physics:hep-th, physics:quant-ph]} \BibitemShut {NoStop}%
\bibitem [{\citenamefont {Pfeuty}(1969)}]{pfeuty1969}%
  \BibitemOpen
  \bibfield  {author} {\bibinfo {author} {\bibfnamefont {P.}~\bibnamefont {Pfeuty}},\ }\bibfield  {title} {\bibinfo {title} {The one-dimensional {{Ising}} model with a transverse field},\ }\href {https://doi.org/10.1016/0003-4916(70)90270-8} {\bibfield  {journal} {\bibinfo  {journal} {Annals of Physics}\ }\textbf {\bibinfo {volume} {57}},\ \bibinfo {pages} {79} (\bibinfo {year} {1969})}\BibitemShut {NoStop}%
\bibitem [{\citenamefont {Fogedby}(1978)}]{fogedby1978}%
  \BibitemOpen
  \bibfield  {author} {\bibinfo {author} {\bibfnamefont {H.~C.}\ \bibnamefont {Fogedby}},\ }\bibfield  {title} {\bibinfo {title} {The ising chain in a skew magnetic field},\ }\href {https://doi.org/10.1088/0022-3719/11/13/025} {\bibfield  {journal} {\bibinfo  {journal} {Journal of Physics C: Solid State Physics}\ }\textbf {\bibinfo {volume} {11}},\ \bibinfo {pages} {2801} (\bibinfo {year} {1978})}\BibitemShut {NoStop}%
\bibitem [{\citenamefont {Ovchinnikov}\ \emph {et~al.}(2003)\citenamefont {Ovchinnikov}, \citenamefont {Dmitriev}, \citenamefont {Krivnov},\ and\ \citenamefont {Cheranovskii}}]{ovchinnikov2003}%
  \BibitemOpen
  \bibfield  {author} {\bibinfo {author} {\bibfnamefont {A.~A.}\ \bibnamefont {Ovchinnikov}}, \bibinfo {author} {\bibfnamefont {D.~V.}\ \bibnamefont {Dmitriev}}, \bibinfo {author} {\bibfnamefont {V.~Y.}\ \bibnamefont {Krivnov}},\ and\ \bibinfo {author} {\bibfnamefont {V.~O.}\ \bibnamefont {Cheranovskii}},\ }\bibfield  {title} {\bibinfo {title} {Antiferromagnetic ising chain in a mixed transverse and longitudinal magnetic field},\ }\href {https://doi.org/10.1103/PhysRevB.68.214406} {\bibfield  {journal} {\bibinfo  {journal} {Phys. Rev. B}\ }\textbf {\bibinfo {volume} {68}},\ \bibinfo {pages} {214406} (\bibinfo {year} {2003})}\BibitemShut {NoStop}%
\bibitem [{\citenamefont {Doyon}(2025)}]{doyon2025}%
  \BibitemOpen
  \bibfield  {author} {\bibinfo {author} {\bibfnamefont {B.}~\bibnamefont {Doyon}},\ }\href {https://doi.org/10.48550/arXiv.2509.15064} {\bibinfo {title} {Twist fields in many-body physics}} (\bibinfo {year} {2025}),\ \Eprint {https://arxiv.org/abs/2509.15064} {arXiv:2509.15064 [math-ph]} \BibitemShut {NoStop}%
\bibitem [{\citenamefont {White}(1992)}]{white1992}%
  \BibitemOpen
  \bibfield  {author} {\bibinfo {author} {\bibfnamefont {S.~R.}\ \bibnamefont {White}},\ }\bibfield  {title} {\bibinfo {title} {Density matrix formulation for quantum renormalization groups},\ }\href {https://doi.org/10.1103/PhysRevLett.69.2863} {\bibfield  {journal} {\bibinfo  {journal} {Phys. Rev. Lett.}\ }\textbf {\bibinfo {volume} {69}},\ \bibinfo {pages} {2863} (\bibinfo {year} {1992})}\BibitemShut {NoStop}%
\bibitem [{\citenamefont {Kormos}\ \emph {et~al.}(2017)\citenamefont {Kormos}, \citenamefont {Collura}, \citenamefont {Tak{\'a}cs},\ and\ \citenamefont {Calabrese}}]{kormos2017}%
  \BibitemOpen
  \bibfield  {author} {\bibinfo {author} {\bibfnamefont {M.}~\bibnamefont {Kormos}}, \bibinfo {author} {\bibfnamefont {M.}~\bibnamefont {Collura}}, \bibinfo {author} {\bibfnamefont {G.}~\bibnamefont {Tak{\'a}cs}},\ and\ \bibinfo {author} {\bibfnamefont {P.}~\bibnamefont {Calabrese}},\ }\bibfield  {title} {\bibinfo {title} {Real-time confinement following a quantum quench to a non-integrable model},\ }\href {https://doi.org/10.1038/nphys3934} {\bibfield  {journal} {\bibinfo  {journal} {Nature Physics}\ }\textbf {\bibinfo {volume} {13}},\ \bibinfo {pages} {246} (\bibinfo {year} {2017})}\BibitemShut {NoStop}%
\bibitem [{\citenamefont {Lagnese}\ \emph {et~al.}(2021)\citenamefont {Lagnese}, \citenamefont {Surace}, \citenamefont {Kormos},\ and\ \citenamefont {Calabrese}}]{lagnese2021}%
  \BibitemOpen
  \bibfield  {author} {\bibinfo {author} {\bibfnamefont {G.}~\bibnamefont {Lagnese}}, \bibinfo {author} {\bibfnamefont {F.~M.}\ \bibnamefont {Surace}}, \bibinfo {author} {\bibfnamefont {M.}~\bibnamefont {Kormos}},\ and\ \bibinfo {author} {\bibfnamefont {P.}~\bibnamefont {Calabrese}},\ }\bibfield  {title} {\bibinfo {title} {False vacuum decay in quantum spin chains},\ }\href {https://doi.org/10.1103/PhysRevB.104.L201106} {\bibfield  {journal} {\bibinfo  {journal} {Physical Review B}\ }\textbf {\bibinfo {volume} {104}},\ \bibinfo {pages} {L201106} (\bibinfo {year} {2021})}\BibitemShut {NoStop}%
\bibitem [{\citenamefont {Villa}\ \emph {et~al.}(2019)\citenamefont {Villa}, \citenamefont {Despres},\ and\ \citenamefont {{Sanchez-Palencia}}}]{villa2019}%
  \BibitemOpen
  \bibfield  {author} {\bibinfo {author} {\bibfnamefont {L.}~\bibnamefont {Villa}}, \bibinfo {author} {\bibfnamefont {J.}~\bibnamefont {Despres}},\ and\ \bibinfo {author} {\bibfnamefont {L.}~\bibnamefont {{Sanchez-Palencia}}},\ }\bibfield  {title} {\bibinfo {title} {Unraveling the excitation spectrum of many-body systems from quantum quenches},\ }\href {https://doi.org/10.1103/PhysRevA.100.063632} {\bibfield  {journal} {\bibinfo  {journal} {Physical Review A}\ }\textbf {\bibinfo {volume} {100}},\ \bibinfo {pages} {063632} (\bibinfo {year} {2019})}\BibitemShut {NoStop}%
\bibitem [{\citenamefont {Villa}\ \emph {et~al.}(2020)\citenamefont {Villa}, \citenamefont {Despres}, \citenamefont {Thomson},\ and\ \citenamefont {{Sanchez-Palencia}}}]{villa2020}%
  \BibitemOpen
  \bibfield  {author} {\bibinfo {author} {\bibfnamefont {L.}~\bibnamefont {Villa}}, \bibinfo {author} {\bibfnamefont {J.}~\bibnamefont {Despres}}, \bibinfo {author} {\bibfnamefont {S.~J.}\ \bibnamefont {Thomson}},\ and\ \bibinfo {author} {\bibfnamefont {L.}~\bibnamefont {{Sanchez-Palencia}}},\ }\bibfield  {title} {\bibinfo {title} {Local quench spectroscopy of many-body quantum systems},\ }\href {https://doi.org/10.1103/PhysRevA.102.033337} {\bibfield  {journal} {\bibinfo  {journal} {Physical Review A}\ }\textbf {\bibinfo {volume} {102}},\ \bibinfo {pages} {033337} (\bibinfo {year} {2020})}\BibitemShut {NoStop}%
\bibitem [{\citenamefont {Chanda}\ \emph {et~al.}(2024)\citenamefont {Chanda}, \citenamefont {Dalmonte}, \citenamefont {Lewenstein}, \citenamefont {Zakrzewski},\ and\ \citenamefont {Tagliacozzo}}]{chanda2024}%
  \BibitemOpen
  \bibfield  {author} {\bibinfo {author} {\bibfnamefont {T.}~\bibnamefont {Chanda}}, \bibinfo {author} {\bibfnamefont {M.}~\bibnamefont {Dalmonte}}, \bibinfo {author} {\bibfnamefont {M.}~\bibnamefont {Lewenstein}}, \bibinfo {author} {\bibfnamefont {J.}~\bibnamefont {Zakrzewski}},\ and\ \bibinfo {author} {\bibfnamefont {L.}~\bibnamefont {Tagliacozzo}},\ }\bibfield  {title} {\bibinfo {title} {Spectral properties of the critical (1+1)-dimensional {{Abelian-Higgs}} model},\ }\href {https://doi.org/10.1103/PhysRevB.109.045103} {\bibfield  {journal} {\bibinfo  {journal} {Physical Review B}\ }\textbf {\bibinfo {volume} {109}},\ \bibinfo {pages} {045103} (\bibinfo {year} {2024})}\BibitemShut {NoStop}%
\bibitem [{\citenamefont {Gross}\ and\ \citenamefont {Bloch}(2017)}]{gross2017}%
  \BibitemOpen
  \bibfield  {author} {\bibinfo {author} {\bibfnamefont {C.}~\bibnamefont {Gross}}\ and\ \bibinfo {author} {\bibfnamefont {I.}~\bibnamefont {Bloch}},\ }\bibfield  {title} {\bibinfo {title} {Quantum simulations with ultracold atoms in optical lattices},\ }\href {https://doi.org/10.1126/science.aal3837} {\bibfield  {journal} {\bibinfo  {journal} {Science}\ }\textbf {\bibinfo {volume} {357}},\ \bibinfo {pages} {995} (\bibinfo {year} {2017})}\BibitemShut {NoStop}%
\bibitem [{\citenamefont {Daley}\ \emph {et~al.}(2022)\citenamefont {Daley}, \citenamefont {Bloch}, \citenamefont {Kokail}, \citenamefont {Flannigan}, \citenamefont {Pearson}, \citenamefont {Troyer},\ and\ \citenamefont {Zoller}}]{daley2022}%
  \BibitemOpen
  \bibfield  {author} {\bibinfo {author} {\bibfnamefont {A.~J.}\ \bibnamefont {Daley}}, \bibinfo {author} {\bibfnamefont {I.}~\bibnamefont {Bloch}}, \bibinfo {author} {\bibfnamefont {C.}~\bibnamefont {Kokail}}, \bibinfo {author} {\bibfnamefont {S.}~\bibnamefont {Flannigan}}, \bibinfo {author} {\bibfnamefont {N.}~\bibnamefont {Pearson}}, \bibinfo {author} {\bibfnamefont {M.}~\bibnamefont {Troyer}},\ and\ \bibinfo {author} {\bibfnamefont {P.}~\bibnamefont {Zoller}},\ }\bibfield  {title} {\bibinfo {title} {Practical quantum advantage in quantum simulation},\ }\href {https://doi.org/10.1038/s41586-022-04940-6} {\bibfield  {journal} {\bibinfo  {journal} {Nature}\ }\textbf {\bibinfo {volume} {607}},\ \bibinfo {pages} {667} (\bibinfo {year} {2022})}\BibitemShut {NoStop}%
\bibitem [{\citenamefont {Bernien}\ \emph {et~al.}(2017)\citenamefont {Bernien}, \citenamefont {Schwartz}, \citenamefont {Keesling}, \citenamefont {Levine}, \citenamefont {Omran}, \citenamefont {Pichler}, \citenamefont {Choi}, \citenamefont {Zibrov}, \citenamefont {Endres}, \citenamefont {Greiner}, \citenamefont {Vuleti{\'c}},\ and\ \citenamefont {Lukin}}]{bernien2017}%
  \BibitemOpen
  \bibfield  {author} {\bibinfo {author} {\bibfnamefont {H.}~\bibnamefont {Bernien}}, \bibinfo {author} {\bibfnamefont {S.}~\bibnamefont {Schwartz}}, \bibinfo {author} {\bibfnamefont {A.}~\bibnamefont {Keesling}}, \bibinfo {author} {\bibfnamefont {H.}~\bibnamefont {Levine}}, \bibinfo {author} {\bibfnamefont {A.}~\bibnamefont {Omran}}, \bibinfo {author} {\bibfnamefont {H.}~\bibnamefont {Pichler}}, \bibinfo {author} {\bibfnamefont {S.}~\bibnamefont {Choi}}, \bibinfo {author} {\bibfnamefont {A.~S.}\ \bibnamefont {Zibrov}}, \bibinfo {author} {\bibfnamefont {M.}~\bibnamefont {Endres}}, \bibinfo {author} {\bibfnamefont {M.}~\bibnamefont {Greiner}}, \bibinfo {author} {\bibfnamefont {V.}~\bibnamefont {Vuleti{\'c}}},\ and\ \bibinfo {author} {\bibfnamefont {M.~D.}\ \bibnamefont {Lukin}},\ }\bibfield  {title} {\bibinfo {title} {Probing many-body dynamics on a 51-atom quantum simulator},\ }\href {https://doi.org/10.1038/nature24622} {\bibfield  {journal} {\bibinfo  {journal} {Nature}\ }\textbf {\bibinfo {volume} {551}},\ \bibinfo {pages} {579} (\bibinfo {year} {2017})}\BibitemShut {NoStop}%
\bibitem [{\citenamefont {Monroe}\ \emph {et~al.}(2021)\citenamefont {Monroe}, \citenamefont {Campbell}, \citenamefont {Duan}, \citenamefont {Gong}, \citenamefont {Gorshkov}, \citenamefont {Hess}, \citenamefont {Islam}, \citenamefont {Kim}, \citenamefont {Linke}, \citenamefont {Pagano}, \citenamefont {Richerme}, \citenamefont {Senko},\ and\ \citenamefont {Yao}}]{monroe2021}%
  \BibitemOpen
  \bibfield  {author} {\bibinfo {author} {\bibfnamefont {C.}~\bibnamefont {Monroe}}, \bibinfo {author} {\bibfnamefont {W.~C.}\ \bibnamefont {Campbell}}, \bibinfo {author} {\bibfnamefont {L.-M.}\ \bibnamefont {Duan}}, \bibinfo {author} {\bibfnamefont {Z.-X.}\ \bibnamefont {Gong}}, \bibinfo {author} {\bibfnamefont {A.~V.}\ \bibnamefont {Gorshkov}}, \bibinfo {author} {\bibfnamefont {P.~W.}\ \bibnamefont {Hess}}, \bibinfo {author} {\bibfnamefont {R.}~\bibnamefont {Islam}}, \bibinfo {author} {\bibfnamefont {K.}~\bibnamefont {Kim}}, \bibinfo {author} {\bibfnamefont {N.~M.}\ \bibnamefont {Linke}}, \bibinfo {author} {\bibfnamefont {G.}~\bibnamefont {Pagano}}, \bibinfo {author} {\bibfnamefont {P.}~\bibnamefont {Richerme}}, \bibinfo {author} {\bibfnamefont {C.}~\bibnamefont {Senko}},\ and\ \bibinfo {author} {\bibfnamefont {N.~Y.}\ \bibnamefont {Yao}},\ }\bibfield  {title} {\bibinfo {title} {Programmable quantum simulations of spin systems with trapped ions},\ }\href {https://doi.org/10.1103/RevModPhys.93.025001} {\bibfield  {journal} {\bibinfo  {journal} {Reviews of Modern Physics}\ }\textbf {\bibinfo {volume} {93}},\ \bibinfo {pages} {025001} (\bibinfo {year} {2021})}\BibitemShut {NoStop}%
\bibitem [{\citenamefont {Rosenberg}\ \emph {et~al.}(2024)\citenamefont {Rosenberg}, \citenamefont {Andersen}, \citenamefont {Samajdar}, \citenamefont {Petukhov}, \citenamefont {Hoke}, \citenamefont {Abanin}, \citenamefont {Bengtsson}, \citenamefont {Drozdov}, \citenamefont {Erickson}, \citenamefont {Klimov}, \citenamefont {Mi}, \citenamefont {Morvan}, \citenamefont {Neeley}, \citenamefont {Neill}, \citenamefont {Acharya}, \citenamefont {Allen}, \citenamefont {Anderson}, \citenamefont {Ansmann}, \citenamefont {Arute}, \citenamefont {Arya}, \citenamefont {Asfaw}, \citenamefont {Atalaya}, \citenamefont {Bardin}, \citenamefont {Bilmes}, \citenamefont {Bortoli}, \citenamefont {Bourassa}, \citenamefont {Bovaird}, \citenamefont {Brill}, \citenamefont {Broughton}, \citenamefont {Buckley}, \citenamefont {Buell}, \citenamefont {Burger}, \citenamefont {Burkett}, \citenamefont {Bushnell}, \citenamefont {Campero}, \citenamefont {Chang}, \citenamefont {Chen}, \citenamefont {Chiaro}, \citenamefont {Chik}, \citenamefont {Cogan}, \citenamefont {Collins}, \citenamefont {Conner}, \citenamefont {Courtney}, \citenamefont {Crook}, \citenamefont {Curtin}, \citenamefont {Debroy}, \citenamefont {Barba}, \citenamefont {Demura}, \citenamefont {Di~Paolo}, \citenamefont {Dunsworth}, \citenamefont {Earle}, \citenamefont {Faoro}, \citenamefont {Farhi}, \citenamefont {Fatemi}, \citenamefont {Ferreira}, \citenamefont {Burgos}, \citenamefont {Forati}, \citenamefont {Fowler}, \citenamefont {Foxen}, \citenamefont {Garcia}, \citenamefont {Genois}, \citenamefont {Giang}, \citenamefont {Gidney}, \citenamefont {Gilboa}, \citenamefont {Giustina}, \citenamefont {Gosula}, \citenamefont {Dau}, \citenamefont {Gross}, \citenamefont {Habegger}, \citenamefont {Hamilton}, \citenamefont {Hansen}, \citenamefont {Harrigan}, \citenamefont {Harrington}, \citenamefont {Heu}, \citenamefont {Hill}, \citenamefont {Hoffmann}, \citenamefont {Hong}, \citenamefont {Huang}, \citenamefont {Huff}, \citenamefont {Huggins}, \citenamefont {Ioffe}, \citenamefont {Isakov}, \citenamefont {Iveland}, \citenamefont {Jeffrey}, \citenamefont {Jiang}, \citenamefont {Jones}, \citenamefont {Juhas}, \citenamefont {Kafri}, \citenamefont {Khattar}, \citenamefont {Khezri}, \citenamefont {Kieferov{\'a}}, \citenamefont {Kim}, \citenamefont {Kitaev}, \citenamefont {Klots}, \citenamefont {Korotkov}, \citenamefont {Kostritsa}, \citenamefont {Kreikebaum}, \citenamefont {Landhuis}, \citenamefont {Laptev}, \citenamefont {Lau}, \citenamefont {Laws}, \citenamefont {Lee}, \citenamefont {Lee}, \citenamefont {Lensky}, \citenamefont {Lester}, \citenamefont {Lill}, \citenamefont {Liu}, \citenamefont {Locharla}, \citenamefont {Mandr{\`a}}, \citenamefont {Martin}, \citenamefont {Martin}, \citenamefont {McClean}, \citenamefont {McEwen}, \citenamefont {Meeks}, \citenamefont {Miao}, \citenamefont {Mieszala}, \citenamefont {Montazeri}, \citenamefont {Movassagh}, \citenamefont {Mruczkiewicz}, \citenamefont {Nersisyan}, \citenamefont {Newman}, \citenamefont {Ng}, \citenamefont {Nguyen}, \citenamefont {Nguyen}, \citenamefont {Niu}, \citenamefont {O'Brien}, \citenamefont {Omonije}, \citenamefont {Opremcak}, \citenamefont {Potter}, \citenamefont {Pryadko}, \citenamefont {Quintana}, \citenamefont {Rhodes}, \citenamefont {Rocque}, \citenamefont {Rubin}, \citenamefont {Saei}, \citenamefont {Sank}, \citenamefont {Sankaragomathi}, \citenamefont {Satzinger}, \citenamefont {Schurkus}, \citenamefont {Schuster}, \citenamefont {Shearn}, \citenamefont {Shorter}, \citenamefont {Shutty}, \citenamefont {Shvarts}, \citenamefont {Sivak}, \citenamefont {Skruzny}, \citenamefont {Smith}, \citenamefont {Somma}, \citenamefont {Sterling}, \citenamefont {Strain}, \citenamefont {Szalay}, \citenamefont {Thor}, \citenamefont {Torres}, \citenamefont {Vidal}, \citenamefont {Villalonga}, \citenamefont {Heidweiller}, \citenamefont {White}, \citenamefont {Woo}, \citenamefont {Xing}, \citenamefont {Yao}, \citenamefont {Yeh}, \citenamefont {Yoo}, \citenamefont {Young}, \citenamefont {Zalcman}, \citenamefont {Zhang}, \citenamefont {Zhu}, \citenamefont {Zobrist}, \citenamefont {Neven}, \citenamefont {Babbush}, \citenamefont {Bacon}, \citenamefont {Boixo}, \citenamefont {Hilton}, \citenamefont {Lucero}, \citenamefont {Megrant}, \citenamefont {Kelly}, \citenamefont {Chen}, \citenamefont {Smelyanskiy}, \citenamefont {Khemani}, \citenamefont {Gopalakrishnan}, \citenamefont {Prosen},\ and\ \citenamefont {Roushan}}]{rosenberg2024}%
  \BibitemOpen
  \bibfield  {author} {\bibinfo {author} {\bibfnamefont {E.}~\bibnamefont {Rosenberg}}, \bibinfo {author} {\bibfnamefont {T.~I.}\ \bibnamefont {Andersen}}, \bibinfo {author} {\bibfnamefont {R.}~\bibnamefont {Samajdar}}, \bibinfo {author} {\bibfnamefont {A.}~\bibnamefont {Petukhov}}, \bibinfo {author} {\bibfnamefont {J.~C.}\ \bibnamefont {Hoke}}, \bibinfo {author} {\bibfnamefont {D.}~\bibnamefont {Abanin}}, \bibinfo {author} {\bibfnamefont {A.}~\bibnamefont {Bengtsson}}, \bibinfo {author} {\bibfnamefont {I.~K.}\ \bibnamefont {Drozdov}}, \bibinfo {author} {\bibfnamefont {C.}~\bibnamefont {Erickson}}, \bibinfo {author} {\bibfnamefont {P.~V.}\ \bibnamefont {Klimov}}, \bibinfo {author} {\bibfnamefont {X.}~\bibnamefont {Mi}}, \bibinfo {author} {\bibfnamefont {A.}~\bibnamefont {Morvan}}, \bibinfo {author} {\bibfnamefont {M.}~\bibnamefont {Neeley}}, \bibinfo {author} {\bibfnamefont {C.}~\bibnamefont {Neill}}, \bibinfo {author} {\bibfnamefont {R.}~\bibnamefont {Acharya}}, \bibinfo {author} {\bibfnamefont {R.}~\bibnamefont {Allen}}, \bibinfo {author} {\bibfnamefont {K.}~\bibnamefont {Anderson}}, \bibinfo {author} {\bibfnamefont {M.}~\bibnamefont {Ansmann}}, \bibinfo {author} {\bibfnamefont {F.}~\bibnamefont {Arute}}, \bibinfo {author} {\bibfnamefont {K.}~\bibnamefont {Arya}}, \bibinfo {author} {\bibfnamefont {A.}~\bibnamefont {Asfaw}}, \bibinfo {author} {\bibfnamefont {J.}~\bibnamefont {Atalaya}}, \bibinfo {author} {\bibfnamefont {J.~C.}\ \bibnamefont {Bardin}}, \bibinfo {author} {\bibfnamefont {A.}~\bibnamefont {Bilmes}}, \bibinfo {author} {\bibfnamefont {G.}~\bibnamefont {Bortoli}}, \bibinfo {author} {\bibfnamefont {A.}~\bibnamefont {Bourassa}}, \bibinfo {author} {\bibfnamefont {J.}~\bibnamefont {Bovaird}}, \bibinfo {author} {\bibfnamefont {L.}~\bibnamefont {Brill}}, \bibinfo {author} {\bibfnamefont {M.}~\bibnamefont {Broughton}}, \bibinfo {author} {\bibfnamefont {B.~B.}\ \bibnamefont {Buckley}}, \bibinfo {author} {\bibfnamefont {D.~A.}\ \bibnamefont {Buell}}, \bibinfo {author} {\bibfnamefont {T.}~\bibnamefont {Burger}}, \bibinfo {author} {\bibfnamefont {B.}~\bibnamefont {Burkett}}, \bibinfo {author} {\bibfnamefont {N.}~\bibnamefont {Bushnell}}, \bibinfo {author} {\bibfnamefont {J.}~\bibnamefont {Campero}}, \bibinfo {author} {\bibfnamefont {H.-S.}\ \bibnamefont {Chang}}, \bibinfo {author} {\bibfnamefont {Z.}~\bibnamefont {Chen}}, \bibinfo {author} {\bibfnamefont {B.}~\bibnamefont {Chiaro}}, \bibinfo {author} {\bibfnamefont {D.}~\bibnamefont {Chik}}, \bibinfo {author} {\bibfnamefont {J.}~\bibnamefont {Cogan}}, \bibinfo {author} {\bibfnamefont {R.}~\bibnamefont {Collins}}, \bibinfo {author} {\bibfnamefont {P.}~\bibnamefont {Conner}}, \bibinfo {author} {\bibfnamefont {W.}~\bibnamefont {Courtney}}, \bibinfo {author} {\bibfnamefont {A.~L.}\ \bibnamefont {Crook}}, \bibinfo {author} {\bibfnamefont {B.}~\bibnamefont {Curtin}}, \bibinfo {author} {\bibfnamefont {D.~M.}\ \bibnamefont {Debroy}}, \bibinfo {author} {\bibfnamefont {A.~D.~T.}\ \bibnamefont {Barba}}, \bibinfo {author} {\bibfnamefont {S.}~\bibnamefont {Demura}}, \bibinfo {author} {\bibfnamefont {A.}~\bibnamefont {Di~Paolo}}, \bibinfo {author} {\bibfnamefont {A.}~\bibnamefont {Dunsworth}}, \bibinfo {author} {\bibfnamefont {C.}~\bibnamefont {Earle}}, \bibinfo {author} {\bibfnamefont {L.}~\bibnamefont {Faoro}}, \bibinfo {author} {\bibfnamefont {E.}~\bibnamefont {Farhi}}, \bibinfo {author} {\bibfnamefont {R.}~\bibnamefont {Fatemi}}, \bibinfo {author} {\bibfnamefont {V.~S.}\ \bibnamefont {Ferreira}}, \bibinfo {author} {\bibfnamefont {L.~F.}\ \bibnamefont {Burgos}}, \bibinfo {author} {\bibfnamefont {E.}~\bibnamefont {Forati}}, \bibinfo {author} {\bibfnamefont {A.~G.}\ \bibnamefont {Fowler}}, \bibinfo {author} {\bibfnamefont {B.}~\bibnamefont {Foxen}}, \bibinfo {author} {\bibfnamefont {G.}~\bibnamefont {Garcia}}, \bibinfo {author} {\bibfnamefont {{\'E}.}~\bibnamefont {Genois}}, \bibinfo {author} {\bibfnamefont {W.}~\bibnamefont {Giang}}, \bibinfo {author} {\bibfnamefont {C.}~\bibnamefont {Gidney}}, \bibinfo {author} {\bibfnamefont {D.}~\bibnamefont {Gilboa}}, \bibinfo {author} {\bibfnamefont {M.}~\bibnamefont {Giustina}}, \bibinfo {author} {\bibfnamefont {R.}~\bibnamefont {Gosula}}, \bibinfo {author} {\bibfnamefont {A.~G.}\ \bibnamefont {Dau}}, \bibinfo {author} {\bibfnamefont {J.~A.}\ \bibnamefont {Gross}}, \bibinfo {author} {\bibfnamefont {S.}~\bibnamefont {Habegger}}, \bibinfo {author} {\bibfnamefont {M.~C.}\ \bibnamefont {Hamilton}}, \bibinfo {author} {\bibfnamefont {M.}~\bibnamefont {Hansen}}, \bibinfo {author} {\bibfnamefont {M.~P.}\ \bibnamefont {Harrigan}}, \bibinfo {author} {\bibfnamefont {S.~D.}\ \bibnamefont {Harrington}}, \bibinfo {author} {\bibfnamefont {P.}~\bibnamefont {Heu}}, \bibinfo {author} {\bibfnamefont {G.}~\bibnamefont {Hill}}, \bibinfo {author} {\bibfnamefont {M.~R.}\ \bibnamefont {Hoffmann}}, \bibinfo {author} {\bibfnamefont {S.}~\bibnamefont {Hong}}, \bibinfo {author} {\bibfnamefont {T.}~\bibnamefont {Huang}}, \bibinfo {author} {\bibfnamefont {A.}~\bibnamefont {Huff}}, \bibinfo {author} {\bibfnamefont {W.~J.}\ \bibnamefont {Huggins}}, \bibinfo {author} {\bibfnamefont {L.~B.}\ \bibnamefont {Ioffe}}, \bibinfo {author} {\bibfnamefont {S.~V.}\ \bibnamefont {Isakov}}, \bibinfo {author} {\bibfnamefont {J.}~\bibnamefont {Iveland}}, \bibinfo {author} {\bibfnamefont {E.}~\bibnamefont {Jeffrey}}, \bibinfo {author} {\bibfnamefont {Z.}~\bibnamefont {Jiang}}, \bibinfo {author} {\bibfnamefont {C.}~\bibnamefont {Jones}}, \bibinfo {author} {\bibfnamefont {P.}~\bibnamefont {Juhas}}, \bibinfo {author} {\bibfnamefont {D.}~\bibnamefont {Kafri}}, \bibinfo {author} {\bibfnamefont {T.}~\bibnamefont {Khattar}}, \bibinfo {author} {\bibfnamefont {M.}~\bibnamefont {Khezri}}, \bibinfo {author} {\bibfnamefont {M.}~\bibnamefont {Kieferov{\'a}}}, \bibinfo {author} {\bibfnamefont {S.}~\bibnamefont {Kim}}, \bibinfo {author} {\bibfnamefont {A.}~\bibnamefont {Kitaev}}, \bibinfo {author} {\bibfnamefont {A.~R.}\ \bibnamefont {Klots}}, \bibinfo {author} {\bibfnamefont {A.~N.}\ \bibnamefont {Korotkov}}, \bibinfo {author} {\bibfnamefont {F.}~\bibnamefont {Kostritsa}}, \bibinfo {author} {\bibfnamefont {J.~M.}\ \bibnamefont {Kreikebaum}}, \bibinfo {author} {\bibfnamefont {D.}~\bibnamefont {Landhuis}}, \bibinfo {author} {\bibfnamefont {P.}~\bibnamefont {Laptev}}, \bibinfo {author} {\bibfnamefont {K.-M.}\ \bibnamefont {Lau}}, \bibinfo {author} {\bibfnamefont {L.}~\bibnamefont {Laws}}, \bibinfo {author} {\bibfnamefont {J.}~\bibnamefont {Lee}}, \bibinfo {author} {\bibfnamefont {K.~W.}\ \bibnamefont {Lee}}, \bibinfo {author} {\bibfnamefont {Y.~D.}\ \bibnamefont {Lensky}}, \bibinfo {author} {\bibfnamefont {B.~J.}\ \bibnamefont {Lester}}, \bibinfo {author} {\bibfnamefont {A.~T.}\ \bibnamefont {Lill}}, \bibinfo {author} {\bibfnamefont {W.}~\bibnamefont {Liu}}, \bibinfo {author} {\bibfnamefont {A.}~\bibnamefont {Locharla}}, \bibinfo {author} {\bibfnamefont {S.}~\bibnamefont {Mandr{\`a}}}, \bibinfo {author} {\bibfnamefont {O.}~\bibnamefont {Martin}}, \bibinfo {author} {\bibfnamefont {S.}~\bibnamefont {Martin}}, \bibinfo {author} {\bibfnamefont {J.~R.}\ \bibnamefont {McClean}}, \bibinfo {author} {\bibfnamefont {M.}~\bibnamefont {McEwen}}, \bibinfo {author} {\bibfnamefont {S.}~\bibnamefont {Meeks}}, \bibinfo {author} {\bibfnamefont {K.~C.}\ \bibnamefont {Miao}}, \bibinfo {author} {\bibfnamefont {A.}~\bibnamefont {Mieszala}}, \bibinfo {author} {\bibfnamefont {S.}~\bibnamefont {Montazeri}}, \bibinfo {author} {\bibfnamefont {R.}~\bibnamefont {Movassagh}}, \bibinfo {author} {\bibfnamefont {W.}~\bibnamefont {Mruczkiewicz}}, \bibinfo {author} {\bibfnamefont {A.}~\bibnamefont {Nersisyan}}, \bibinfo {author} {\bibfnamefont {M.}~\bibnamefont {Newman}}, \bibinfo {author} {\bibfnamefont {J.~H.}\ \bibnamefont {Ng}}, \bibinfo {author} {\bibfnamefont {A.}~\bibnamefont {Nguyen}}, \bibinfo {author} {\bibfnamefont {M.}~\bibnamefont {Nguyen}}, \bibinfo {author} {\bibfnamefont {M.~Y.}\ \bibnamefont {Niu}}, \bibinfo {author} {\bibfnamefont {T.~E.}\ \bibnamefont {O'Brien}}, \bibinfo {author} {\bibfnamefont {S.}~\bibnamefont {Omonije}}, \bibinfo {author} {\bibfnamefont {A.}~\bibnamefont {Opremcak}}, \bibinfo {author} {\bibfnamefont {R.}~\bibnamefont {Potter}}, \bibinfo {author} {\bibfnamefont {L.~P.}\ \bibnamefont {Pryadko}}, \bibinfo {author} {\bibfnamefont {C.}~\bibnamefont {Quintana}}, \bibinfo {author} {\bibfnamefont {D.~M.}\ \bibnamefont {Rhodes}}, \bibinfo {author} {\bibfnamefont {C.}~\bibnamefont {Rocque}}, \bibinfo {author} {\bibfnamefont {N.~C.}\ \bibnamefont {Rubin}}, \bibinfo {author} {\bibfnamefont {N.}~\bibnamefont {Saei}}, \bibinfo {author} {\bibfnamefont {D.}~\bibnamefont {Sank}}, \bibinfo {author} {\bibfnamefont {K.}~\bibnamefont {Sankaragomathi}}, \bibinfo {author} {\bibfnamefont {K.~J.}\ \bibnamefont {Satzinger}}, \bibinfo {author} {\bibfnamefont {H.~F.}\ \bibnamefont {Schurkus}}, \bibinfo {author} {\bibfnamefont {C.}~\bibnamefont {Schuster}}, \bibinfo {author} {\bibfnamefont {M.~J.}\ \bibnamefont {Shearn}}, \bibinfo {author} {\bibfnamefont {A.}~\bibnamefont {Shorter}}, \bibinfo {author} {\bibfnamefont {N.}~\bibnamefont {Shutty}}, \bibinfo {author} {\bibfnamefont {V.}~\bibnamefont {Shvarts}}, \bibinfo {author} {\bibfnamefont {V.}~\bibnamefont {Sivak}}, \bibinfo {author} {\bibfnamefont {J.}~\bibnamefont {Skruzny}}, \bibinfo {author} {\bibfnamefont {W.~C.}\ \bibnamefont {Smith}}, \bibinfo {author} {\bibfnamefont {R.~D.}\ \bibnamefont {Somma}}, \bibinfo {author} {\bibfnamefont {G.}~\bibnamefont {Sterling}}, \bibinfo {author} {\bibfnamefont {D.}~\bibnamefont {Strain}}, \bibinfo {author} {\bibfnamefont {M.}~\bibnamefont {Szalay}}, \bibinfo {author} {\bibfnamefont {D.}~\bibnamefont {Thor}}, \bibinfo {author} {\bibfnamefont {A.}~\bibnamefont {Torres}}, \bibinfo {author} {\bibfnamefont {G.}~\bibnamefont {Vidal}}, \bibinfo {author} {\bibfnamefont {B.}~\bibnamefont {Villalonga}}, \bibinfo {author} {\bibfnamefont {C.~V.}\ \bibnamefont {Heidweiller}}, \bibinfo {author} {\bibfnamefont {T.}~\bibnamefont {White}},
  \bibinfo {author} {\bibfnamefont {B.~W.~K.}\ \bibnamefont {Woo}}, \bibinfo {author} {\bibfnamefont {C.}~\bibnamefont {Xing}}, \bibinfo {author} {\bibfnamefont {Z.~J.}\ \bibnamefont {Yao}}, \bibinfo {author} {\bibfnamefont {P.}~\bibnamefont {Yeh}}, \bibinfo {author} {\bibfnamefont {J.}~\bibnamefont {Yoo}}, \bibinfo {author} {\bibfnamefont {G.}~\bibnamefont {Young}}, \bibinfo {author} {\bibfnamefont {A.}~\bibnamefont {Zalcman}}, \bibinfo {author} {\bibfnamefont {Y.}~\bibnamefont {Zhang}}, \bibinfo {author} {\bibfnamefont {N.}~\bibnamefont {Zhu}}, \bibinfo {author} {\bibfnamefont {N.}~\bibnamefont {Zobrist}}, \bibinfo {author} {\bibfnamefont {H.}~\bibnamefont {Neven}}, \bibinfo {author} {\bibfnamefont {R.}~\bibnamefont {Babbush}}, \bibinfo {author} {\bibfnamefont {D.}~\bibnamefont {Bacon}}, \bibinfo {author} {\bibfnamefont {S.}~\bibnamefont {Boixo}}, \bibinfo {author} {\bibfnamefont {J.}~\bibnamefont {Hilton}}, \bibinfo {author} {\bibfnamefont {E.}~\bibnamefont {Lucero}}, \bibinfo {author} {\bibfnamefont {A.}~\bibnamefont {Megrant}}, \bibinfo {author} {\bibfnamefont {J.}~\bibnamefont {Kelly}}, \bibinfo {author} {\bibfnamefont {Y.}~\bibnamefont {Chen}}, \bibinfo {author} {\bibfnamefont {V.}~\bibnamefont {Smelyanskiy}}, \bibinfo {author} {\bibfnamefont {V.}~\bibnamefont {Khemani}}, \bibinfo {author} {\bibfnamefont {S.}~\bibnamefont {Gopalakrishnan}}, \bibinfo {author} {\bibfnamefont {T.}~\bibnamefont {Prosen}},\ and\ \bibinfo {author} {\bibfnamefont {P.}~\bibnamefont {Roushan}},\ }\bibfield  {title} {\bibinfo {title} {Dynamics of magnetization at infinite temperature in a {{Heisenberg}} spin chain},\ }\href {https://doi.org/10.1126/science.adi7877} {\bibfield  {journal} {\bibinfo  {journal} {Science}\ }\textbf {\bibinfo {volume} {384}},\ \bibinfo {pages} {48} (\bibinfo {year} {2024})}\BibitemShut {NoStop}%
\bibitem [{\citenamefont {Wienand}\ \emph {et~al.}(2024)\citenamefont {Wienand}, \citenamefont {Karch}, \citenamefont {Impertro}, \citenamefont {Schweizer}, \citenamefont {McCulloch}, \citenamefont {Vasseur}, \citenamefont {Gopalakrishnan}, \citenamefont {Aidelsburger},\ and\ \citenamefont {Bloch}}]{wienand2024}%
  \BibitemOpen
  \bibfield  {author} {\bibinfo {author} {\bibfnamefont {J.~F.}\ \bibnamefont {Wienand}}, \bibinfo {author} {\bibfnamefont {S.}~\bibnamefont {Karch}}, \bibinfo {author} {\bibfnamefont {A.}~\bibnamefont {Impertro}}, \bibinfo {author} {\bibfnamefont {C.}~\bibnamefont {Schweizer}}, \bibinfo {author} {\bibfnamefont {E.}~\bibnamefont {McCulloch}}, \bibinfo {author} {\bibfnamefont {R.}~\bibnamefont {Vasseur}}, \bibinfo {author} {\bibfnamefont {S.}~\bibnamefont {Gopalakrishnan}}, \bibinfo {author} {\bibfnamefont {M.}~\bibnamefont {Aidelsburger}},\ and\ \bibinfo {author} {\bibfnamefont {I.}~\bibnamefont {Bloch}},\ }\bibfield  {title} {\bibinfo {title} {Emergence of fluctuating hydrodynamics in chaotic quantum systems},\ }\href {https://doi.org/10.1038/s41567-024-02611-z} {\bibfield  {journal} {\bibinfo  {journal} {Nature Physics}\ ,\ \bibinfo {pages} {1}} (\bibinfo {year} {2024})}\BibitemShut {NoStop}%
\bibitem [{\citenamefont {Su}\ \emph {et~al.}(2023)\citenamefont {Su}, \citenamefont {Sun}, \citenamefont {Hudomal}, \citenamefont {Desaules}, \citenamefont {Zhou}, \citenamefont {Yang}, \citenamefont {Halimeh}, \citenamefont {Yuan}, \citenamefont {Papi{\'c}},\ and\ \citenamefont {Pan}}]{su2023}%
  \BibitemOpen
  \bibfield  {author} {\bibinfo {author} {\bibfnamefont {G.-X.}\ \bibnamefont {Su}}, \bibinfo {author} {\bibfnamefont {H.}~\bibnamefont {Sun}}, \bibinfo {author} {\bibfnamefont {A.}~\bibnamefont {Hudomal}}, \bibinfo {author} {\bibfnamefont {J.-Y.}\ \bibnamefont {Desaules}}, \bibinfo {author} {\bibfnamefont {Z.-Y.}\ \bibnamefont {Zhou}}, \bibinfo {author} {\bibfnamefont {B.}~\bibnamefont {Yang}}, \bibinfo {author} {\bibfnamefont {J.~C.}\ \bibnamefont {Halimeh}}, \bibinfo {author} {\bibfnamefont {Z.-S.}\ \bibnamefont {Yuan}}, \bibinfo {author} {\bibfnamefont {Z.}~\bibnamefont {Papi{\'c}}},\ and\ \bibinfo {author} {\bibfnamefont {J.-W.}\ \bibnamefont {Pan}},\ }\bibfield  {title} {\bibinfo {title} {Observation of unconventional many-body scarring in a quantum simulator},\ }\href {https://doi.org/10.1103/PhysRevResearch.5.023010} {\bibfield  {journal} {\bibinfo  {journal} {Physical Review Research}\ }\textbf {\bibinfo {volume} {5}},\ \bibinfo {pages} {023010} (\bibinfo {year} {2023})},\ \Eprint {https://arxiv.org/abs/2201.00821} {arXiv:2201.00821} \BibitemShut {NoStop}%
\bibitem [{\citenamefont {Sachdev}\ \emph {et~al.}(2002)\citenamefont {Sachdev}, \citenamefont {Sengupta},\ and\ \citenamefont {Girvin}}]{sachdev2002}%
  \BibitemOpen
  \bibfield  {author} {\bibinfo {author} {\bibfnamefont {S.}~\bibnamefont {Sachdev}}, \bibinfo {author} {\bibfnamefont {K.}~\bibnamefont {Sengupta}},\ and\ \bibinfo {author} {\bibfnamefont {S.~M.}\ \bibnamefont {Girvin}},\ }\bibfield  {title} {\bibinfo {title} {Mott insulators in strong electric fields},\ }\href {https://doi.org/10.1103/PhysRevB.66.075128} {\bibfield  {journal} {\bibinfo  {journal} {Physical Review B}\ }\textbf {\bibinfo {volume} {66}},\ \bibinfo {pages} {075128} (\bibinfo {year} {2002})}\BibitemShut {NoStop}%
\bibitem [{\citenamefont {Schweizer}\ \emph {et~al.}(2019)\citenamefont {Schweizer}, \citenamefont {Grusdt}, \citenamefont {Berngruber}, \citenamefont {Barbiero}, \citenamefont {Demler}, \citenamefont {Goldman}, \citenamefont {Bloch},\ and\ \citenamefont {Aidelsburger}}]{schweizer2019}%
  \BibitemOpen
  \bibfield  {author} {\bibinfo {author} {\bibfnamefont {C.}~\bibnamefont {Schweizer}}, \bibinfo {author} {\bibfnamefont {F.}~\bibnamefont {Grusdt}}, \bibinfo {author} {\bibfnamefont {M.}~\bibnamefont {Berngruber}}, \bibinfo {author} {\bibfnamefont {L.}~\bibnamefont {Barbiero}}, \bibinfo {author} {\bibfnamefont {E.}~\bibnamefont {Demler}}, \bibinfo {author} {\bibfnamefont {N.}~\bibnamefont {Goldman}}, \bibinfo {author} {\bibfnamefont {I.}~\bibnamefont {Bloch}},\ and\ \bibinfo {author} {\bibfnamefont {M.}~\bibnamefont {Aidelsburger}},\ }\bibfield  {title} {\bibinfo {title} {Floquet approach to {{$\mathbb{Z}$2}} lattice gauge theories with ultracold atoms in optical lattices},\ }\href {https://doi.org/10.1038/s41567-019-0649-7} {\bibfield  {journal} {\bibinfo  {journal} {Nature Physics}\ }\textbf {\bibinfo {volume} {15}},\ \bibinfo {pages} {1168} (\bibinfo {year} {2019})}\BibitemShut {NoStop}%
\bibitem [{\citenamefont {Yang}\ \emph {et~al.}(2020)\citenamefont {Yang}, \citenamefont {Sun}, \citenamefont {Ott}, \citenamefont {Wang}, \citenamefont {Zache}, \citenamefont {Halimeh}, \citenamefont {Yuan}, \citenamefont {Hauke},\ and\ \citenamefont {Pan}}]{yang2020}%
  \BibitemOpen
  \bibfield  {author} {\bibinfo {author} {\bibfnamefont {B.}~\bibnamefont {Yang}}, \bibinfo {author} {\bibfnamefont {H.}~\bibnamefont {Sun}}, \bibinfo {author} {\bibfnamefont {R.}~\bibnamefont {Ott}}, \bibinfo {author} {\bibfnamefont {H.-Y.}\ \bibnamefont {Wang}}, \bibinfo {author} {\bibfnamefont {T.~V.}\ \bibnamefont {Zache}}, \bibinfo {author} {\bibfnamefont {J.~C.}\ \bibnamefont {Halimeh}}, \bibinfo {author} {\bibfnamefont {Z.-S.}\ \bibnamefont {Yuan}}, \bibinfo {author} {\bibfnamefont {P.}~\bibnamefont {Hauke}},\ and\ \bibinfo {author} {\bibfnamefont {J.-W.}\ \bibnamefont {Pan}},\ }\bibfield  {title} {\bibinfo {title} {Observation of gauge invariance in a 71-site {{Bose}}--{{Hubbard}} quantum simulator},\ }\href {https://doi.org/10.1038/s41586-020-2910-8} {\bibfield  {journal} {\bibinfo  {journal} {Nature}\ }\textbf {\bibinfo {volume} {587}},\ \bibinfo {pages} {392} (\bibinfo {year} {2020})}\BibitemShut {NoStop}%
\bibitem [{\citenamefont {Viermann}\ \emph {et~al.}(2022)\citenamefont {Viermann}, \citenamefont {Sparn}, \citenamefont {Liebster}, \citenamefont {Hans}, \citenamefont {Kath}, \citenamefont {{Parra-L{\'o}pez}}, \citenamefont {{Tolosa-Sime{\'o}n}}, \citenamefont {{S{\'a}nchez-Kuntz}}, \citenamefont {Haas}, \citenamefont {Strobel}, \citenamefont {Floerchinger},\ and\ \citenamefont {Oberthaler}}]{viermann2022}%
  \BibitemOpen
  \bibfield  {author} {\bibinfo {author} {\bibfnamefont {C.}~\bibnamefont {Viermann}}, \bibinfo {author} {\bibfnamefont {M.}~\bibnamefont {Sparn}}, \bibinfo {author} {\bibfnamefont {N.}~\bibnamefont {Liebster}}, \bibinfo {author} {\bibfnamefont {M.}~\bibnamefont {Hans}}, \bibinfo {author} {\bibfnamefont {E.}~\bibnamefont {Kath}}, \bibinfo {author} {\bibfnamefont {{\'A}.}~\bibnamefont {{Parra-L{\'o}pez}}}, \bibinfo {author} {\bibfnamefont {M.}~\bibnamefont {{Tolosa-Sime{\'o}n}}}, \bibinfo {author} {\bibfnamefont {N.}~\bibnamefont {{S{\'a}nchez-Kuntz}}}, \bibinfo {author} {\bibfnamefont {T.}~\bibnamefont {Haas}}, \bibinfo {author} {\bibfnamefont {H.}~\bibnamefont {Strobel}}, \bibinfo {author} {\bibfnamefont {S.}~\bibnamefont {Floerchinger}},\ and\ \bibinfo {author} {\bibfnamefont {M.~K.}\ \bibnamefont {Oberthaler}},\ }\bibfield  {title} {\bibinfo {title} {Quantum field simulator for dynamics in curved spacetime},\ }\href {https://doi.org/10.1038/s41586-022-05313-9} {\bibfield  {journal} {\bibinfo  {journal} {Nature}\ }\textbf {\bibinfo {volume} {611}},\ \bibinfo {pages} {260} (\bibinfo {year} {2022})}\BibitemShut {NoStop}%
\bibitem [{\citenamefont {Guo}\ \emph {et~al.}(2024{\natexlab{b}})\citenamefont {Guo}, \citenamefont {Wu}, \citenamefont {Ye}, \citenamefont {Zhang}, \citenamefont {Lian}, \citenamefont {Yao}, \citenamefont {Wang}, \citenamefont {Yan}, \citenamefont {Yi}, \citenamefont {Xu}, \citenamefont {Li}, \citenamefont {Hou}, \citenamefont {Xu}, \citenamefont {Guo}, \citenamefont {Zhang}, \citenamefont {Qi}, \citenamefont {Zhou}, \citenamefont {He},\ and\ \citenamefont {Duan}}]{guo2024b}%
  \BibitemOpen
  \bibfield  {author} {\bibinfo {author} {\bibfnamefont {S.-A.}\ \bibnamefont {Guo}}, \bibinfo {author} {\bibfnamefont {Y.-K.}\ \bibnamefont {Wu}}, \bibinfo {author} {\bibfnamefont {J.}~\bibnamefont {Ye}}, \bibinfo {author} {\bibfnamefont {L.}~\bibnamefont {Zhang}}, \bibinfo {author} {\bibfnamefont {W.-Q.}\ \bibnamefont {Lian}}, \bibinfo {author} {\bibfnamefont {R.}~\bibnamefont {Yao}}, \bibinfo {author} {\bibfnamefont {Y.}~\bibnamefont {Wang}}, \bibinfo {author} {\bibfnamefont {R.-Y.}\ \bibnamefont {Yan}}, \bibinfo {author} {\bibfnamefont {Y.-J.}\ \bibnamefont {Yi}}, \bibinfo {author} {\bibfnamefont {Y.-L.}\ \bibnamefont {Xu}}, \bibinfo {author} {\bibfnamefont {B.-W.}\ \bibnamefont {Li}}, \bibinfo {author} {\bibfnamefont {Y.-H.}\ \bibnamefont {Hou}}, \bibinfo {author} {\bibfnamefont {Y.-Z.}\ \bibnamefont {Xu}}, \bibinfo {author} {\bibfnamefont {W.-X.}\ \bibnamefont {Guo}}, \bibinfo {author} {\bibfnamefont {C.}~\bibnamefont {Zhang}}, \bibinfo {author} {\bibfnamefont {B.-X.}\ \bibnamefont {Qi}}, \bibinfo {author} {\bibfnamefont {Z.-C.}\ \bibnamefont {Zhou}}, \bibinfo {author} {\bibfnamefont {L.}~\bibnamefont {He}},\ and\ \bibinfo {author} {\bibfnamefont {L.-M.}\ \bibnamefont {Duan}},\ }\bibfield  {title} {\bibinfo {title} {A site-resolved two-dimensional quantum simulator with hundreds of trapped ions},\ }\href {https://doi.org/10.1038/s41586-024-07459-0} {\bibfield  {journal} {\bibinfo  {journal} {Nature}\ }\textbf {\bibinfo {volume} {630}},\ \bibinfo {pages} {613} (\bibinfo {year} {2024}{\natexlab{b}})}\BibitemShut {NoStop}%
\bibitem [{\citenamefont {Qiao}\ \emph {et~al.}(2024)\citenamefont {Qiao}, \citenamefont {Cai}, \citenamefont {Wang}, \citenamefont {Du}, \citenamefont {Jin}, \citenamefont {Chen}, \citenamefont {Wang}, \citenamefont {Luan}, \citenamefont {Gao}, \citenamefont {Sun}, \citenamefont {Tian}, \citenamefont {Zhang},\ and\ \citenamefont {Kim}}]{qiao2024}%
  \BibitemOpen
  \bibfield  {author} {\bibinfo {author} {\bibfnamefont {M.}~\bibnamefont {Qiao}}, \bibinfo {author} {\bibfnamefont {Z.}~\bibnamefont {Cai}}, \bibinfo {author} {\bibfnamefont {Y.}~\bibnamefont {Wang}}, \bibinfo {author} {\bibfnamefont {B.}~\bibnamefont {Du}}, \bibinfo {author} {\bibfnamefont {N.}~\bibnamefont {Jin}}, \bibinfo {author} {\bibfnamefont {W.}~\bibnamefont {Chen}}, \bibinfo {author} {\bibfnamefont {P.}~\bibnamefont {Wang}}, \bibinfo {author} {\bibfnamefont {C.}~\bibnamefont {Luan}}, \bibinfo {author} {\bibfnamefont {E.}~\bibnamefont {Gao}}, \bibinfo {author} {\bibfnamefont {X.}~\bibnamefont {Sun}}, \bibinfo {author} {\bibfnamefont {H.}~\bibnamefont {Tian}}, \bibinfo {author} {\bibfnamefont {J.}~\bibnamefont {Zhang}},\ and\ \bibinfo {author} {\bibfnamefont {K.}~\bibnamefont {Kim}},\ }\bibfield  {title} {\bibinfo {title} {Tunable quantum simulation of spin models with a two-dimensional ion crystal},\ }\href {https://doi.org/10.1038/s41567-023-02378-9} {\bibfield  {journal} {\bibinfo  {journal} {Nature Physics}\ }\textbf {\bibinfo {volume} {20}},\ \bibinfo {pages} {623} (\bibinfo {year} {2024})}\BibitemShut {NoStop}%
\bibitem [{\citenamefont {Barredo}\ \emph {et~al.}(2016)\citenamefont {Barredo}, \citenamefont {De~L{\'e}s{\'e}leuc}, \citenamefont {Lienhard}, \citenamefont {Lahaye},\ and\ \citenamefont {Browaeys}}]{barredo2016}%
  \BibitemOpen
  \bibfield  {author} {\bibinfo {author} {\bibfnamefont {D.}~\bibnamefont {Barredo}}, \bibinfo {author} {\bibfnamefont {S.}~\bibnamefont {De~L{\'e}s{\'e}leuc}}, \bibinfo {author} {\bibfnamefont {V.}~\bibnamefont {Lienhard}}, \bibinfo {author} {\bibfnamefont {T.}~\bibnamefont {Lahaye}},\ and\ \bibinfo {author} {\bibfnamefont {A.}~\bibnamefont {Browaeys}},\ }\bibfield  {title} {\bibinfo {title} {An atom-by-atom assembler of defect-free arbitrary two-dimensional atomic arrays},\ }\href {https://doi.org/10.1126/science.aah3778} {\bibfield  {journal} {\bibinfo  {journal} {Science}\ }\textbf {\bibinfo {volume} {354}},\ \bibinfo {pages} {1021} (\bibinfo {year} {2016})}\BibitemShut {NoStop}%
\bibitem [{\citenamefont {Labuhn}\ \emph {et~al.}(2016)\citenamefont {Labuhn}, \citenamefont {Barredo}, \citenamefont {Ravets}, \citenamefont {{de L{\'e}s{\'e}leuc}}, \citenamefont {Macr{\`i}}, \citenamefont {Lahaye},\ and\ \citenamefont {Browaeys}}]{labuhn2016}%
  \BibitemOpen
  \bibfield  {author} {\bibinfo {author} {\bibfnamefont {H.}~\bibnamefont {Labuhn}}, \bibinfo {author} {\bibfnamefont {D.}~\bibnamefont {Barredo}}, \bibinfo {author} {\bibfnamefont {S.}~\bibnamefont {Ravets}}, \bibinfo {author} {\bibfnamefont {S.}~\bibnamefont {{de L{\'e}s{\'e}leuc}}}, \bibinfo {author} {\bibfnamefont {T.}~\bibnamefont {Macr{\`i}}}, \bibinfo {author} {\bibfnamefont {T.}~\bibnamefont {Lahaye}},\ and\ \bibinfo {author} {\bibfnamefont {A.}~\bibnamefont {Browaeys}},\ }\bibfield  {title} {\bibinfo {title} {Tunable two-dimensional arrays of single {{Rydberg}} atoms for realizing quantum {{Ising}} models},\ }\href {https://doi.org/10.1038/nature18274} {\bibfield  {journal} {\bibinfo  {journal} {Nature}\ }\textbf {\bibinfo {volume} {534}},\ \bibinfo {pages} {667} (\bibinfo {year} {2016})}\BibitemShut {NoStop}%
\bibitem [{\citenamefont {Scholl}\ \emph {et~al.}(2021)\citenamefont {Scholl}, \citenamefont {Schuler}, \citenamefont {Williams}, \citenamefont {Eberharter}, \citenamefont {Barredo}, \citenamefont {Schymik}, \citenamefont {Lienhard}, \citenamefont {Henry}, \citenamefont {Lang}, \citenamefont {Lahaye}, \citenamefont {L{\"a}uchli},\ and\ \citenamefont {Browaeys}}]{scholl2021a}%
  \BibitemOpen
  \bibfield  {author} {\bibinfo {author} {\bibfnamefont {P.}~\bibnamefont {Scholl}}, \bibinfo {author} {\bibfnamefont {M.}~\bibnamefont {Schuler}}, \bibinfo {author} {\bibfnamefont {H.~J.}\ \bibnamefont {Williams}}, \bibinfo {author} {\bibfnamefont {A.~A.}\ \bibnamefont {Eberharter}}, \bibinfo {author} {\bibfnamefont {D.}~\bibnamefont {Barredo}}, \bibinfo {author} {\bibfnamefont {K.-N.}\ \bibnamefont {Schymik}}, \bibinfo {author} {\bibfnamefont {V.}~\bibnamefont {Lienhard}}, \bibinfo {author} {\bibfnamefont {L.-P.}\ \bibnamefont {Henry}}, \bibinfo {author} {\bibfnamefont {T.~C.}\ \bibnamefont {Lang}}, \bibinfo {author} {\bibfnamefont {T.}~\bibnamefont {Lahaye}}, \bibinfo {author} {\bibfnamefont {A.~M.}\ \bibnamefont {L{\"a}uchli}},\ and\ \bibinfo {author} {\bibfnamefont {A.}~\bibnamefont {Browaeys}},\ }\bibfield  {title} {\bibinfo {title} {Quantum simulation of {{2D}} antiferromagnets with hundreds of {{Rydberg}} atoms},\ }\href {https://doi.org/10.1038/s41586-021-03585-1} {\bibfield  {journal} {\bibinfo  {journal} {Nature}\ }\textbf {\bibinfo {volume} {595}},\ \bibinfo {pages} {233} (\bibinfo {year} {2021})}\BibitemShut {NoStop}%
\bibitem [{\citenamefont {Semeghini}\ \emph {et~al.}(2021)\citenamefont {Semeghini}, \citenamefont {Levine}, \citenamefont {Keesling}, \citenamefont {Ebadi}, \citenamefont {Wang}, \citenamefont {Bluvstein}, \citenamefont {Verresen}, \citenamefont {Pichler}, \citenamefont {Kalinowski}, \citenamefont {Samajdar}, \citenamefont {Omran}, \citenamefont {Sachdev}, \citenamefont {Vishwanath}, \citenamefont {Greiner}, \citenamefont {Vuleti{\'c}},\ and\ \citenamefont {Lukin}}]{semeghini2021}%
  \BibitemOpen
  \bibfield  {author} {\bibinfo {author} {\bibfnamefont {G.}~\bibnamefont {Semeghini}}, \bibinfo {author} {\bibfnamefont {H.}~\bibnamefont {Levine}}, \bibinfo {author} {\bibfnamefont {A.}~\bibnamefont {Keesling}}, \bibinfo {author} {\bibfnamefont {S.}~\bibnamefont {Ebadi}}, \bibinfo {author} {\bibfnamefont {T.~T.}\ \bibnamefont {Wang}}, \bibinfo {author} {\bibfnamefont {D.}~\bibnamefont {Bluvstein}}, \bibinfo {author} {\bibfnamefont {R.}~\bibnamefont {Verresen}}, \bibinfo {author} {\bibfnamefont {H.}~\bibnamefont {Pichler}}, \bibinfo {author} {\bibfnamefont {M.}~\bibnamefont {Kalinowski}}, \bibinfo {author} {\bibfnamefont {R.}~\bibnamefont {Samajdar}}, \bibinfo {author} {\bibfnamefont {A.}~\bibnamefont {Omran}}, \bibinfo {author} {\bibfnamefont {S.}~\bibnamefont {Sachdev}}, \bibinfo {author} {\bibfnamefont {A.}~\bibnamefont {Vishwanath}}, \bibinfo {author} {\bibfnamefont {M.}~\bibnamefont {Greiner}}, \bibinfo {author} {\bibfnamefont {V.}~\bibnamefont {Vuleti{\'c}}},\ and\ \bibinfo {author} {\bibfnamefont {M.~D.}\ \bibnamefont {Lukin}},\ }\bibfield  {title} {\bibinfo {title} {Probing topological spin liquids on a programmable quantum simulator},\ }\href {https://doi.org/10.1126/science.abi8794} {\bibfield  {journal} {\bibinfo  {journal} {Science (New York, N.Y.)}\ }\textbf {\bibinfo {volume} {374}},\ \bibinfo {pages} {1242} (\bibinfo {year} {2021})}\BibitemShut {NoStop}%
\bibitem [{\citenamefont {Milekhin}\ \emph {et~al.}(2025)\citenamefont {Milekhin}, \citenamefont {Adamska},\ and\ \citenamefont {Preskill}}]{milekhin2025a}%
  \BibitemOpen
  \bibfield  {author} {\bibinfo {author} {\bibfnamefont {A.}~\bibnamefont {Milekhin}}, \bibinfo {author} {\bibfnamefont {Z.}~\bibnamefont {Adamska}},\ and\ \bibinfo {author} {\bibfnamefont {J.}~\bibnamefont {Preskill}},\ }\href {https://doi.org/10.48550/arXiv.2502.12240} {\bibinfo {title} {Observable and computable entanglement in time}} (\bibinfo {year} {2025}),\ \Eprint {https://arxiv.org/abs/2502.12240} {arXiv:2502.12240} \BibitemShut {NoStop}%
\end{thebibliography}%

\begin{appendices}
\section{Probing the quench dynamics with different operators}
\label{app:diif_op}
\begin{figure*}

  \includegraphics[width=\linewidth]{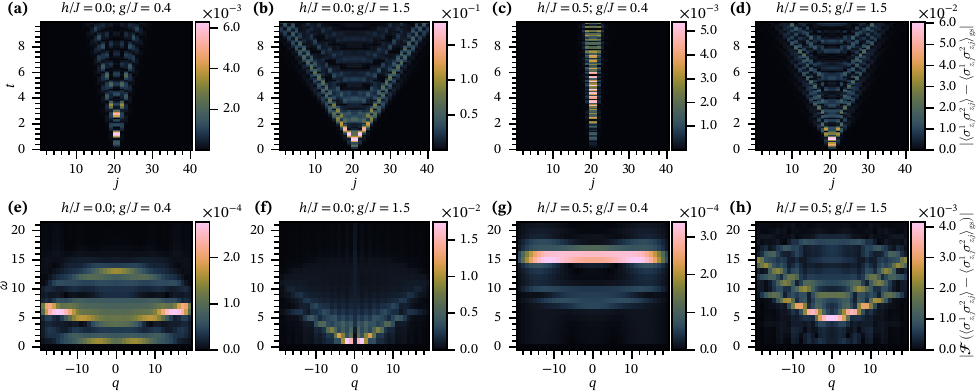}\\
  \caption{\label{fig:fourier_ana_zz}%
  Temporal purities of the operator $\sigma_z$ for systems made by \(N=40\) spins in the ground state of the Ising model \cref{eq:ham}. Panels (a)-(d) show their space-time profile (with time measured in units of inverse energy (\(1/J\))), while the corresponding Fourier transforms obtained via Eq.~\eqref{eq:fourier} are in (e)-(h). Integrable cases: (a) and (e) lie in the ferromagnetic phase, \(g/J=0.4\) and \(h/J=0\);  (b) and (f) in the paramagnetic phase  \(g/J=1.5\) and \(h/J=0\).  Non-integrable cases (c) and (g) show the data for  \(g/J=0.4\) and \(h/J=0.5\); (d) and (h) correspond to  \(g/J=1.5\) and \(h/J=0.5\). 
  }
\end{figure*}

In~\cref{sect:pump-probe} we have made a connection between the measurement of generalized temporal R\'enyi entropies and pump-probe experiments. The geometric quench we design initially excites some energy eigenmodes of the system. Given that we deal with gapped systems, we expect that only the low energy excitations are involved in the subsequent dynamics.
After letting the system evolve for some time, we then probe the dynamics. Here different operators could in principle probe different branches of excitations. However, in general, we expect that most local operators will initially have uniform matrix elements on the system eigenmodes, before relaxing to the diagonal ensemble, as dictated by the eigenstate thermalization hypothesis.

As a consequence, in generic systems, we expect that the response we measure using different operators should not drastically differ: by measuring few local operators we should thus have a consistent picture of the low energy dispersion relation.

Here we will check this assumption on the Ising model by comparing the results of the quench extracted from different operators and showing that the relevant features  are visible can be extracted by studying any of them.
 Concretely, the feature we want to characterize in   ~\cref{fig:fourier_ana_1} and ~\cref{fig:fourier_ana_2} is the presence/absence of a gap in the dispersion relation extracted from space-time Fourier transform of the connected correlation functions.

 In this section, we show that the gap in Fourier space is a generic feature of the system and appears in all the operators we have been characterizing. In~\cref{fig:fourier_ana_zz} we study the behavior of the connected correlator of $\langle \sigma_z^1\sigma_z^2\rangle(t)-\langle \sigma_z^1\sigma_z^2\rangle_{gs}$.
 Panels (a)-(d) show the real space behavior, while panels (e)-(h) show its Fourier transform, obtained following~\cref{eq:fourier}. The Fourier transform of the quench dynamics in integrable cases, panels (e) and (f), shows a $q=0$ and $\omega=0$ mode, as expected from the discussion in~\cref{sec:footprint-integrability}. On the other hand, in the non-integrable cases, panels (f) and (h), the $q=0$ and $\omega=0$ mode is absent, in fact the spectra is gapped. 

 Comparing Fig.~\ref{fig:fourier_ana_1}(e-h) and Fig.~\ref{fig:fourier_ana_zz}(e-h), one can clearly see that there are some details that are different for different probing operators. For example, Fig.~\ref{fig:fourier_ana_1}(h) displays two branches with their endpoints near $\omega = 15$, on the other hand, in Fig.~\ref{fig:fourier_ana_zz}(h) we can still see these branches but additionally, a new branch is shown with its minimum around $\omega = 5$. Similar details change in the rest of the cases, but the presence or absence  of the soft mode is a stable feature of the system and in our simulation only seems to depend on whether the system is governed by a dynamics in the  integrable or  non-integrable regime. In the integrable regime, also, given the replicated nature of the operator, we can only probe $Z^2$ even operators, and thus the odd sector could be gapped. However, given we are interested in the presence/absence of a gapless mode, it is sufficient to observe it in the even sector.

\section{Finite-size and boundary effects}

\begin{figure*}
  \includegraphics[width=\linewidth]{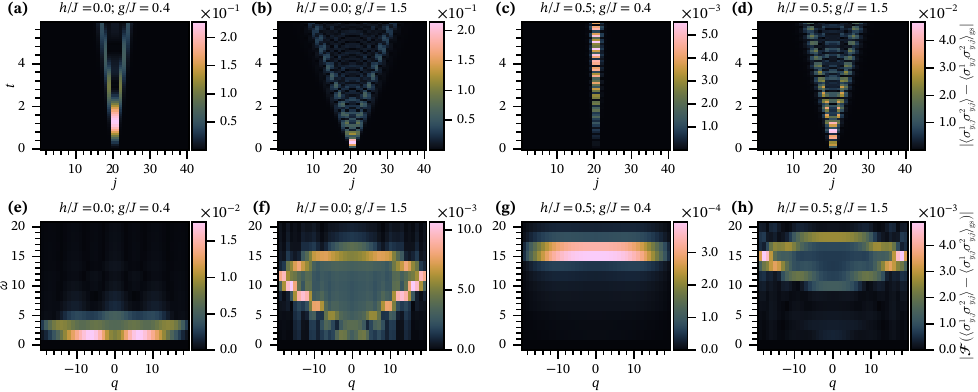}\\
  \caption{\label{fig:fourier_ana_yy6}%
  Temporal purities for systems made by \(N=40\) spins in the ground state of the Ising model \cref{eq:ham}. Panels (a)-(d) show their space time profile (with time measured in units of inverse energy (\(1/J\))), the corresponding Fourier transforms obtained via Eq.~\eqref{eq:fourier} are shown in (e)-(h), with frequency measured in units of energy (\(J\)). Integrable cases: (a) and (e) lie in the ferromagnetic phase, \(g/J=0.4\) and \(h/J=0\);  (b) and (f) in the paramagnetic phase  \(g/J=1.5\) and \(h/J=0\).  Non-integrable cases (c) and (g) show the data for  \(g/J=0.4\) and \(h/J=0.5\); (d) and (h) correspond to  \(g/J=1.5\) and \(h/J=0.5\). The  bright regions in (e) and (f) close to $\omega=0$ and $q=0$ unveil the appearance of soft modes in the integrable cases which are absent in (g) and (h). The systems are evolved up to a final time $TJ=6$.}
\end{figure*}

Up to this point we have assumed that the results we have presented, obtained by characterizing finite-size systems, are actually correctly reproducing the infinite-system results given that the space-time signals we extract  have a clear causal cone and, in most cases, for the times we have been analyzing, such cone has not yet reached the boundaries of the system.

It is clear that a truly gapless mode can be observed only in the thermodynamic limit, though in all cases we have considered the soft mode we observe is sufficiently separated from the gap of the system with a that we can safely assume it would become gapless in the $N\to \infty$ limit.

Such an assumption is confirmed  in ~\cref{sec:footprint-integrability}, where in ~\cref{fig:fourier_ana_2}(c) we have been able to consider the thermodynamic limit,  ($N\rightarrow\infty$), and have actually observed the emergence of the gapless excitation in the  integrable dynamics.

In~\cref{fig:fourier_ana_1,fig:fourier_ana_zz}(c)  the light-cone of the quench reaches the boundary of the system at time $t \approx 9$, and thus this gives us the opportunity to address how  the finite-size effects affect our ability to predict the presence/ absence of a  gap.

We thus study the importance of the finite-size systems from another point of view by performing the Fourier study for different times, where the light-cone has not reached the boundaries of the system (that is, for $TJ=6$ and $TJ=8$), to see whether the system displays the same features as in the case of~\cref{fig:fourier_ana_1}.

\begin{figure*}
  \includegraphics[width=\linewidth]{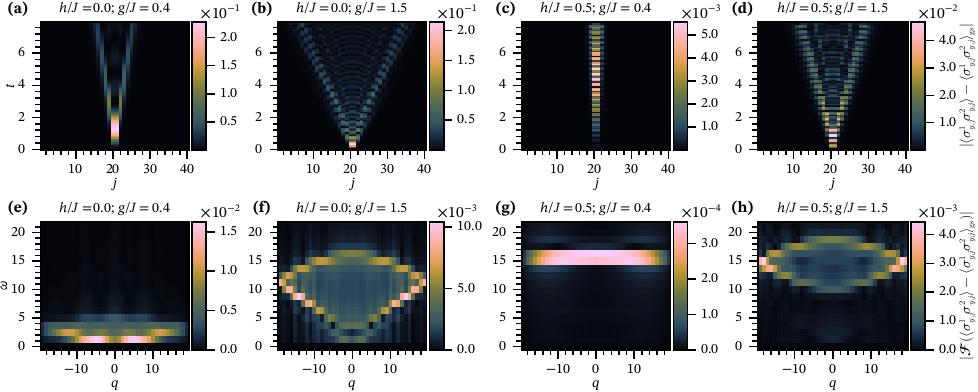}\\
  \caption{\label{fig:fourier_ana_yy8}%
  Temporal purities for systems made by \(N=40\) spins in the ground state of the Ising model \cref{eq:ham}. Panels (a)-(d) show their space time profile (with time measured in units of inverse energy (\(1/J\))) and the corresponding Fourier transforms obtained via Eq.~\eqref{eq:fourier} are in (e)-(h), with frequency measured in units of energy (\(J\)). Integrable cases: (a) and (e) lie in the ferromagnetic phase, \(g/J=0.4\) and \(h/J=0\);  (b) and (f) in the paramagnetic phase  \(g/J=1.5\) and \(h/J=0\).  Non-integrable cases (c) and (g) show the data for  \(g/J=0.4\) and \(h/J=0.5\); (d) and (h) correspond to  \(g/J=1.5\) and \(h/J=0.5\). The  bright regions in (e) and (f) close to $\omega=0$ and $q=0$ unveil the appearance of soft modes in the integrable cases which are absent in (g) and (h). The systems are evolved up to a final time $TJ=8$.}
\end{figure*}

For simplicity, we focus only on the behavior of $\langle \sigma_y^1\sigma_y^2\rangle(t)-\langle \sigma_y^1\sigma_y^2\rangle_{gs}$ and its corresponding Fourier transform, since, we have discussed in~\cref{app:diif_op}, the presence/absence of the gap does not depend on the operator used to probe the system. ~\cref{fig:fourier_ana_yy6,fig:fourier_ana_yy8} displays the same simulations as~\cref{fig:fourier_ana_1}, but stopped before reaching the boundary, at $TJ=6$ and $TJ=8$ respectively. The light-cone visible in the space-time profile of the operator now never reaches the boundaries of the system.

By performing the  Fourier transformation for the simulations reaching $TJ=6$,~\cref{fig:fourier_ana_yy6}(e-h), we have a very good  agreement with~\cref{fig:fourier_ana_1}, despite the worst frequency resolution caused by the shorter time series. The order of magnitude and the shape of the transforms are consistent among the figures and the discrepancies are compatible with being caused by the broadening of the dispersion relation induced by the difference in resolution between the two figures, $\delta \omega =\frac{2\pi}{6}-\frac{2\pi}{8}\simeq 10^{-1}$. The same results are obtained for relatively longer times, $TJ=8$,~\cref{fig:fourier_ana_yy8}(e-h) where the shape and the order of magnitude of the Fourier transform of the excitation $\langle \sigma_y^1\sigma_y^2\rangle(t)-\langle \sigma_y^1\sigma_y^2\rangle_{gs}$ are once more fully compatible with those extracted  in~\cref{fig:fourier_ana_1}(e-h) once the lower resolution is taken in consideration.

We can thus conclude that the results in~\cref{fig:fourier_ana_1} are not polluted with finite-size effects and in combination with~\ref{fig:fourier_ana_2}(c) we conclude that the soft mode of finite size system becomes truly gapless in the thermodynamic limit for the integrable dynamics.
\end{appendices}

\end{document}